\documentstyle[pre,aps,multicol,epsf]{revtex}

\draft

\begin{document}

\title{Surface Critical Behavior in Systems with Non-Equilibrium
Phase Transitions}

\author{Martin Howard,$^{1,}$\cite{MH}$^,$\cite{MH1} 
	Per Fr\"ojdh,$^{2,}$\cite{PF} and 
	Kent B{\ae}kgaard Lauritsen$^{3,4,}$\cite{KBL}}

\address{$^1$Department of Physics, Virginia Tech, Blacksburg, VA
        24061-0435, USA}
\address{$^2$Department of Physics, Stockholm University, Box 6730,
        S-113 85 Stockholm, Sweden}
\address{$^3$Niels Bohr Institute, Center for Chaos and Turbulence
        Studies, DK--2100 Copenhagen, Denmark}
\address{$^4$ Danish Meteorological Institute, Atmosphere Ionosphere
        Research Division, DK--2100 Copenhagen, Denmark}
\date{\today}

\maketitle

\begin{abstract}
We study the surface critical behavior of branching-annihilating
random walks with an even number of offspring (BARW) and directed
percolation (DP) using a variety of
theoretical techniques. Above the upper critical dimensions $d_c$,
with $d_c=4$ (DP) and $d_c=2$ (BARW),
we use mean field theory to analyze the surface phase
diagrams using the standard classification into ordinary, special,
surface, and extraordinary transitions.
For the case of BARW, at or below the upper critical dimension $d\leq
d_c$, we use field theoretic methods to study the effects of
fluctuations. As in the bulk, the field theory suffers from technical
difficulties associated with the presence of a second critical
dimension. However, we are still able to analyze the phase diagrams
for BARW in $d=1,2$, which turn out to be very different from their
mean field analog. Furthermore, for the case
of BARW only (and {\it not} for DP), we find two independent surface
$\beta_1$ exponents in $d=1$, arising from two distinct definitions of
the order parameter. Using an exact duality transformation on a
lattice BARW model in $d=1$, we uncover 
a relationship between these two surface $\beta_1$ 
exponents at the ordinary and special transitions. Many of our
predictions are supported using Monte-Carlo simulations of two
different models belonging to the BARW universality class.

\end{abstract}


\begin{multicols}{2}
\narrowtext


\section{Introduction}

The study of surface critical behavior in equilibrium statistical
mechanics has established the importance of boundaries in critical
systems and their impact on scaling and universality
\cite{review_surf}. 
Quantities measured close to the surface can scale differently than in
the bulk and can possess distinct critical surface exponents. 
Depending on the boundary conditions, various surface
universality classes are possible each with different values for
the surface exponents. In this paper we will be interested in the 
surface critical behavior of certain {\it dynamic} systems,
which possess a non-equilibrium phase transition from an active 
into an absorbing state from which the system cannot escape.

The most prominent example of a system with an absorbing state
is {\it directed percolation} (DP). It describes the directed growth
of a cluster governed by a growth probability $p$ of its fundamental
constituents. For probabilities below a critical value,
$p < p_c$, the cluster will die after a finite time, which means
that the system gets trapped in the vacuum -- the unique empty state.
On the other hand, for high enough growth probabilities
$p > p_c$, there is a finite probability that the cluster will always
remain active. Exactly at $p=p_c$, there is a
critical phase transition from the active into the absorbing state
\cite{kinzel}.
A whole range of other systems possessing a phase transition from a
non-trivial active phase into a unique absorbing state fall into this
universality class. Some examples include epidemics, chemical
reactions, catalysis, and the contact process (see \cite{dickman} and
references therein) .

During the last few years, however, studies have also been carried out
for systems with absorbing states which do not belong to the DP class.
For instance, the model of {\it branching-annihilating 
random walks} with an even number of offspring (BARW) exhibits quite
different behavior \cite{tt,jensen:1994,cardy-tauber} and defines a
separate universality class. Other models in this class (at least in
$d=1$) include certain probabilistic
cellular automata \cite{automata}, monomer-dimer models
\cite{monomerdimer,hwang-etal:1998,inf_DP_DI}, non-equilibrium kinetic
Ising models 
\cite{menyhard-odor}, and generalized DP with two absorbing states
(DP2) \cite{haye}. These models escape from the DP universality class
by possessing an extra conservation law or symmetry: for the BARW
model, a ``parity'' conservation of the total number of particles
modulo $2$; for the other models, an underlying symmetry between their
absorbing states. 

In the present paper we study the impact of surfaces on
the critical behavior of the DP and BARW models.
Previous work has concentrated on
surface effects in DP using field theoretic methods
\cite{janssen-etal,dp-wall-edge}, Monte-Carlo simulations in $d=1,2$ 
\cite{dp-wall-edge,lauritsen-etal,menezes-moukarzel}, the Density
Matrix Renormalization Group in $d=1$ \cite{nancy}, and series
expansions in $d=1$ \cite{essam-etal:1996,newjensen}.
Relations between surface DP and local persistence probabilities have
been explored in \cite{hayehari}. Work has also been performed on
active, but slanted, walls in DP, which give rise 
to a ``curtain'' of activity whose width is given by an
angle-dependent correction to bulk DP \cite{seattle}. 
Critical surface effects
for a model in the BARW universality class were first studied in
\cite{ourprl}. In this paper we will build on this earlier work by
presenting a unified picture of surface critical behavior of both BARW
and DP. After summarizing the main details
of DP and BARW in the bulk, we present a
comprehensive analysis of the surface critical 
behavior of both models using mean field theory. This involves the
usual 
classification into ordinary, special, surface, and extraordinary
transitions. However, below their respective upper critical
dimensions, fluctuation effects become 
important in both models, and this leads to the breakdown of mean
field theory. In order to understand this fluctuation regime we
employ a variety of theoretical techniques. First, we construct a
phenomenological scaling theory which is able to describe the various
surface universality classes. This scaling theory for BARW can then,
to some extent, be justified using field theoretic techniques (a field
theory for surface DP has already been presented in
\cite{janssen-etal}). However, the BARW field theory suffers from
technical problems associated with the presence of a second critical
dimension, which means that the interesting $d=1$ regime
cannot be accessed satisfactorily. Nevertheless some results can
be derived field theoretically which we put together to draw up a
$1+1$ dimensional
surface phase diagram. This phase diagram displays many differences
from its mean field 
analog. In addition, using exact techniques involving
a mapping to a quantum spin Hamiltonian, we have been able to
establish an exact duality transformation for a lattice BARW model in
$d=1$. We find that this links together two of the boundary
phase transitions in $d=1$ in a non-trivial way (as was suggested in 
\cite{ourprl}). We have also performed
extensive Monte-Carlo simulations for BARW and DP2 which support many
of our theoretical conclusions. 

The paper is organized as follows: in Section~\ref{BARWbulk} we
briefly introduce the bulk DP and BARW models. Then in
Section~\ref{DPsurface} we consider the surface behavior of DP, where
we present an extensive mean field analysis. We also summarize
details of the fluctuation regime for $d<d_c=4$. In
Sections~\ref{mfbarw} and \ref{d=1}, we give the phase diagram for
surface BARW 
in mean field theory and in $d=1$ respectively. These results can be
contained within a scaling theory presented in
Section~\ref{corrbarw}. We then discuss in Section~\ref{fieldth} how
this analysis can be partially justified using field theoretic
methods. In Section~\ref{exres}, we give some exact results for
$d=1$. Our theoretical analysis is then backed up using computer
simulations of the lattice models 
introduced in Section~\ref{dpnbarw}. Details of these simulations are
presented in Section~\ref{numres}.
Finally in Section~\ref{conc} we round off with some conclusions.

\section{Bulk DP and BARW}
\label{BARWbulk}

\begin{figure}[htb]
\centerline{\hbox{
\epsfxsize=1.5in
\epsfbox{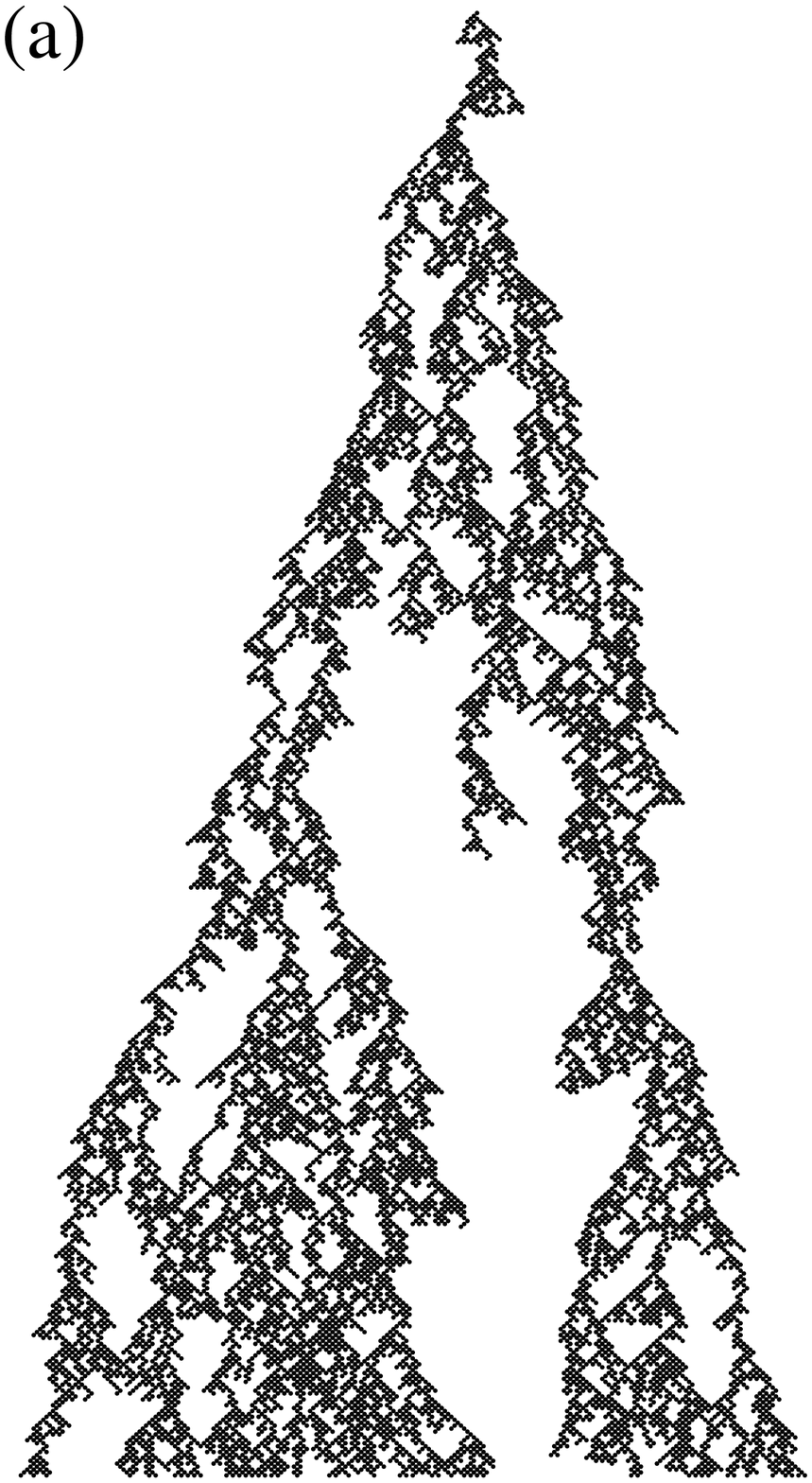}}
\epsfxsize=1.5in
\epsfbox{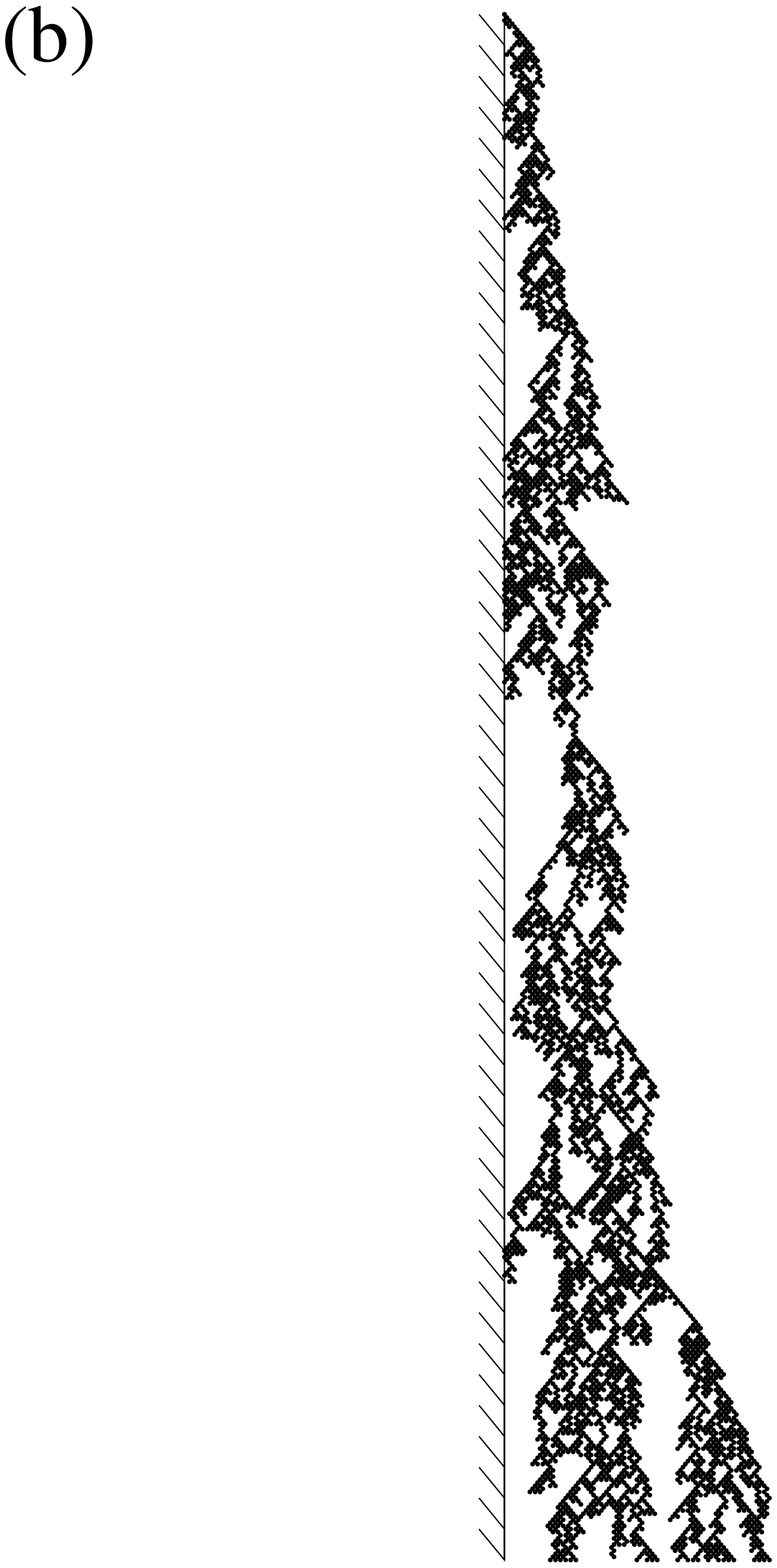}}
\vspace*{5mm}
\caption{DP clusters grown from a single seed (a) in the bulk and (b)
  next to a wall.} 
\label{dpfig}
\end{figure}

We begin by briefly reviewing the definitions of the DP and BARW
models. The update rules for bond DP in $d+1$ dimensions on a tilted
square lattice are easily
defined: for each site at time $t$, form bonds with probability $p$
to the neighboring sites at time $t+1$ \cite{kinzel}. An example of a
cluster grown from a single seed according to these rules is shown in
Figure~\ref{dpfig}a. 

For growth probabilities below a certain threshold such a process
will eventually die out, whereas for higher values there is a finite
probability of survival, which means that the system is in the
active state \cite{grassberger-torre}. 
As is well known \cite{cardy-sugar,janssen1981,sundermeyer}, various
reaction-diffusion models also fall into the DP universality
class. The 
simplest of these is defined by the following reaction scheme for a
single species of diffusing particles:
\begin{eqnarray}
& A \to A+A & {\rm ~with~rate~}\sigma \nonumber \\
& A+A \to A & {\rm ~with~rate~}\lambda \label{dpreacs} \\
& A \to \emptyset & {\rm ~with~rate~}\mu , \nonumber
\end{eqnarray} 
where, in the corresponding lattice model, we allow for
multiple (bosonic) occupancy of any given site.

The second system which we will analyze in detail is the BARW
model \cite{tt,jensen:1994,cardy-tauber}. This is defined again
by a (bosonic) particle model, with the following reaction processes
\begin{eqnarray}
& A \to (m+1)A & {\rm ~with~rate~}\sigma_m \nonumber \\
& A+A\to\emptyset & {\rm ~with~rate~}\lambda . \label{barwreacs}
\end{eqnarray}
For $m$ odd, the above model is known to belong to the DP universality
class, however for $m$ even we have a new universality class. Unless
otherwise specified when we refer to the BARW model we will be
referring to the even $m$ case.

\begin{figure}
\begin{center}
\leavevmode
\vbox{
\epsfxsize=3in
\epsffile{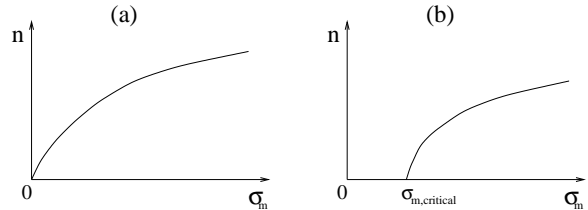}}
\end{center}
\caption{Schematic bulk behavior for BARW of the density $n$ as a function of
the branching rate $\sigma_m$ for (a) $d\geq 2$ and (b) $d=1$.} 
\label{pbulk}
\end{figure}

The growth of both BARW and DP clusters in the bulk close to criticality 
can be summarized by a set
of independent exponents. A natural choice is to consider 
$\nu_\perp$ and $\nu_\parallel$ which describe the divergence of the 
correlation lengths in space, 
$\xi_\perp \sim |\Delta|^{-\nu_\perp}$, and time $\xi_\parallel
\sim |\Delta|^{-\nu_\parallel}$. Here the parameter $\Delta$
describes the deviation from the critical point (in mean field theory 
$\Delta = \mu-\sigma$ for DP, but $\Delta=-m\sigma_m$ for BARW). 
We also need the order parameter exponent $\beta$, which can 
be defined in two {\it a priori} different ways: it is either governed by 
the percolation probability (the probability that a cluster grown from a 
finite seed never dies),
\begin{equation}
\label{P_bulk}
         P(t\to\infty,\Delta) \sim |\Delta|^{\beta_{\rm seed}}, \qquad
         \Delta < 0, 
\end{equation}
or by the coarse-grained density of active sites in the steady state,
\begin{equation}
\label{n(Delta)}
         n(\Delta) \sim |\Delta|^{\beta_{\rm dens}}, \qquad \Delta < 0.
\end{equation}
When $\Delta<0$ the system is said to be in an
{\it active} state, whereas for $\Delta=0$ the system is
{\it critical} (with an algebraically decaying density), and for
$\Delta>0$ (if applicable) the system is {\it inactive}
(with an exponentially decaying density) \cite{diffprl}.
For the case of DP, it is known that $\beta$ is unique: 
$\beta_{\rm seed} = \beta_{\rm dens}$ in any dimension, both above and
below the upper critical dimension $d_c=4$. 
This follows from field theoretic considerations 
\cite{grassberger-torre,cardy-sugar} and has been verified by extensive 
numerical work. The relation also holds for BARW in $1+1$ dimensions,
a result first suggested by numerics and now backed up by an exact
duality mapping \cite{schutz2}. However, this exponent equality
is certainly not always true: if we consider the BARW mean-field regime
valid for spatial dimensions $d>d_c=2$, then
the system is in a critical inactive state only for a zero branching
rate, where the density decays away as a power law.
However, any non-zero branching rate results in an active state, with
a non-zero steady state density (see Figure~\ref{pbulk}a)
\cite{cardy-tauber}. This density (\ref{n(Delta)}) approaches zero
continuously (as the branching rate is reduced towards zero)
with the mean-field exponent $\beta_{\rm dens}=1$. 
Nevertheless, for $d>2$, the survival probability 
(\ref{P_bulk}) of a particle cluster will be finite for {\it any\/} 
value of the branching rate, implying that 
$\beta_{\rm seed}=0$ in mean-field theory. This result follows from 
the non-recurrence of random walks in $d>2$. 

Field theoretically, DP is believed to be satisfactorily 
understood---the appropriate field theory (sometimes called Reggeon Field 
Theory) \cite{cardy-sugar,sundermeyer} is
well under control and the exponents have been computed to two loop
order in an $\epsilon=4-d$ expansion \cite{janssen1981}.
However, for the case of BARW, a description of the $1+1$ dimensional
case poses considerable difficulties for the field theory
\cite{cardy-tauber}. 
These stem from the presence of two critical dimensions: $d_c=2$ (above
which mean-field theory applies) and $d_c' \approx 4/3$. For $d>d_c'$
the behavior of Figure~\ref{pbulk}a holds, i.e. an active state
results for 
{\it any} non-zero value of the branching $\sigma_m$, whereas for
$d<d_c'$ the system is only active for $\sigma_m>\sigma_{m,{\rm
critical}}$, as shown in Figure~\ref{pbulk}b
\cite{cardy-tauber}. This means 
that the physical spatial dimension $d=1$ cannot be
accessed using epsilon expansions down from the upper critical
dimension  $d_c=2$. Furthermore, for the $\sigma_m<\sigma_{m,{\rm
critical}}$ region, the system is {\it not} inactive (in the 
sense of an exponentially decaying density). Instead this entire phase
is controlled by the annihilation fixed point of the $A+A\to\emptyset$
process, where the density decays away as a power law. Hence this
phase should rather be considered as still being critical.

Despite the problems associated with BARW for $d<d_c'$,
we can still put forward a general scaling theory for DP and
BARW, valid both above and below their critical dimensions. However,
we must retain a possible distinction between $\beta_{\rm seed}$ and
$\beta_{\rm dens}$. For example, 
the average lifetime $\langle t\rangle$ of finite clusters can be
derived from the scaling form for the survival probability
\begin{equation}
\label{survbulk}
P(t,\Delta)=|\Delta|^{\beta_{\rm seed}}\varphi(t/\xi_{\parallel}) .
\end{equation} 
We then find $\langle t \rangle \sim |\Delta|^{-\tau}$, where 
$\tau = \nu_\parallel - \beta_{\rm seed}$. The appropriate scaling
form for the density $n({\bf x},t)$, given that the cluster was started
at ${\bf x}={\bf 0},t=0$, is 
\begin{equation}
\label{densbulk}
n(x,t,\Delta)=|\Delta|^{\beta_{\rm seed}+\beta_{\rm
dens}}f(x/\xi_{\perp},t/\xi_{\parallel}) .
\end{equation}
Notice that rotational symmetry about the seeding point ${\bf x}=0$
implies that
the spatial coordinates enter the scaling function only as
$x=|{\bf x}|$, the distance from the seeding point.
Using the expression (\ref{densbulk}) we see that the average mass of
finite clusters, 
$\langle s \rangle \sim |\Delta|^{-\gamma}$, is related to the other
exponents via the following hyperscaling relation:
\begin{equation}
        \label{gen_bulk_hyperscaling}
        \nu_{\parallel} +d\nu_{\perp} 
                = \beta_{\rm seed} + \beta_{\rm dens} +\gamma .
\end{equation} 
Note that (\ref{gen_bulk_hyperscaling}) is consistent with the
distinct upper critical dimensions for BARW and DP. Using the above
mean-field values for BARW and $\nu_\perp=1/2$, $\nu_\parallel=1$, and 
$\gamma=1$, we verify $d_c = 2$. 
In contrast, for DP one has the mean-field exponents $\beta_{\rm
dens}=\beta_{\rm seed} = 1$ and $d_c = 4$.

\section{Surface DP}
\label{DPsurface}

We now briefly review the surface critical behavior of DP and
indicate how the above relations and exponents are modified in a
semi-infinite geometry, where we place a wall at 
$x_\perp=0$ [${\bf x}=({\bf x_{\parallel}},x_{\perp})$, with the $\perp$
and $\parallel$ directions being relative to the wall]. An example of
such a cluster grown close to a wall is shown in Figure~\ref{dpfig}b.

\begin{figure}
\begin{center}
\leavevmode
\vbox{
\epsfxsize=2.5in
\epsffile{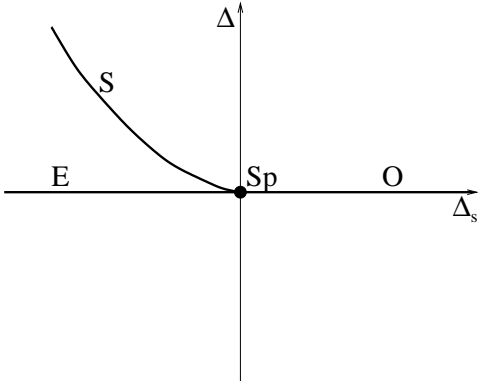}}
\end{center}
\caption{Schematic mean field phase diagram for surface DP. See text
for an explanation of the labeling.} 
\label{dpps}
\end{figure}

A schematic phase diagram for surface DP is shown in Figure~\ref{dpps}
(see \cite{janssen-etal}), where $\Delta_s$ is the deviation of
the surface from criticality. In Figure~\ref{dpps}, the labeling
conforms to 
the standard nomenclature of surface critical phenomena: O stands for
the ordinary transition (bulk critical, surface inactive); Sp is for
the special Transition (bulk and surface both critical); S is for the
surface transition (surface critical, bulk inactive); and finally
E stands for the extraordinary transition (surface active,
bulk critical). 

The bulk exponents are, of course, unchanged by the presence of a
surface and, furthermore, one can show that the correlation
length exponents on the surface are also the same as in the bulk. 
Hence, except at the special transition, one finds just {\it one} extra
exponent: the surface density exponent $\beta_{1,{\rm dens}}$. This
is defined from the steady-state density at the wall. For example, at
the ordinary transition, we have
\begin{equation}
	n(x_\perp=0,\Delta) \sim |\Delta|^{\beta^{\rm O}_{1,{\rm dens}}},
	                    \qquad  \Delta<0. 
\end{equation}
On the other hand, at the multicritical special transition,
one finds {\it two} independent surface exponents -- a new
surface density exponent, $\beta^{\rm Sp}_{1,{\rm dens}}$, and a
crossover exponent $\phi_1$. 
In principle one could also allow for a second
type of surface $\beta_1$ exponent, one defined from a
survival probability for clusters started on the wall. For example, at
the ordinary transition, we would have 
\begin{equation}
	P_1(t\to\infty,\Delta) \sim |\Delta|^{\beta^{\rm O}_{1,{\rm seed}}},
                               \qquad \Delta<0.
\end{equation}
However, the surface exponents here show a similar pattern to their
bulk counterparts and fulfill
$\beta^{\rm O}_{1, \rm seed} = \beta^{\rm O}_{1,{\rm dens}}
=\beta_1^{\rm O}$, 
as can be shown by a field-theoretic derivation of an appropriate 
correlation function \cite{dp-wall-edge}. This kind of equality
should also hold for the $\beta_1$ exponents at the special and
surface transitions.

Numerically, the exponents at the ordinary transition
have been measured very accurately using series expansions (for $d=1$)
\cite{essam-etal:1996,newjensen}, and Monte-Carlo simulations (for $d=1,2$) 
\cite{dp-wall-edge,lauritsen-etal,menezes-moukarzel}.
However, there has been no numerical work to
date on any of the other possible transitions on the boundary.

\subsection{Mean Field Theory}
\label{DPmft}

Although a
considerable amount of work has already been performed on surface DP
\cite{janssen-etal,dp-wall-edge,lauritsen-etal,essam-etal:1996}, a
comprehensive mean 
field analysis has been lacking. The purpose of this section is to
provide such an analysis, and in the process we will derive several new
results. The equation describing mean field DP with a surface is
\begin{equation}
	\label{dpmfteq}
	\partial_t n = D\nabla^2 n -\Delta n -\lambda n^2 ,
\end{equation}
with the boundary condition 
\begin{equation}
	D \partial_{x_{\perp}}n|_{x_{\perp}=0}= \Delta_s n|_{x_{\perp}=0}.
						\label{eq:BCgeneral}
\end{equation}
Here the variable $\Delta=\mu-\sigma$ is the difference
between the rates for the $A\to\emptyset$ and $A\to A+A$
processes. Similarly we have the surface variable $\Delta_s$,
and the bulk
quadratic term is due to the reaction $A+A\to A$. Note that a surface
$A+A\to A$ reaction does not have to be included, as it is an
irrelevant process in the renormalization group (RG) sense
\cite{janssen-etal}. From the above equation (\ref{dpmfteq}), the bulk
mean field exponents can easily be
computed: $\nu_{\parallel}=1$, $\nu_{\perp}=1/2$, and
$\beta=1$. Furthermore, with the inclusion of a boundary, we see that
the correlation length exponents are unchanged at the wall but the
surface $\beta_1$ exponents are altered. If we are interested in the mean
field steady state, then we can replace (\ref{dpmfteq}) with
\begin{equation}
	\label{dpmftred}
	D n'' -\Delta n -\lambda n^2 =0 ,
\end{equation}
where $n''\equiv d^2n/dx_{\perp}^2$. The appropriate boundary
condition~(\ref{eq:BCgeneral}) is given by 
\begin{equation}
	D n'_s = \Delta_s n_s, 
                              \label{eq:BC}
\end{equation}
where $n_s=n|_{x_{\perp}=0}$, and
$n'_s=dn/dx_{\perp}|_{x_{\perp}=0}$. Multiplying (\ref{dpmftred}) by
$n'$ and integrating, we have
\begin{equation}
\label{int1}
{1\over 2}Dn'^2-{1\over 2}\Delta n^2-{1\over 3}\lambda n^3 +C =0 ,
\end{equation}
where $C$ is a constant of integration. Using the bulk results $n'=0$,
and $n=(-\Delta)/\lambda$ for $\Delta<0$, or $n=0$ for $\Delta>0$,
we have 
\begin{eqnarray}
\label{int2a}
& {\Delta_s n_s\over D} = -\left[{\lambda\over D}\right]^{1/2}\left(n_s-
{|\Delta|\over\lambda}\right)\left({2\over 3}n_s+{|\Delta|\over
3\lambda}\right)^{1/2} & [\Delta<0]
\hspace{-.1in} \\
\label{int2b}
& {\Delta_s n_s\over D} = -\left[{\lambda\over
D}\right]^{1/2}n_s\left({2\over 
3}n_s+{\Delta\over\lambda}\right)^{1/2} & [\Delta>0]
\hspace{-.1in} 
\end{eqnarray}
where we have also used the boundary condition~(\ref{eq:BC}).

$\bullet$ Ordinary Transition. Consider the case where $\Delta_s>0$ and
$\Delta\to 0^-$. In that case we expect $n=|\Delta|/\lambda\gg n_s$,
and thus Eq.~(\ref{int2a}) yields $n_s\propto|\Delta|^{3/2}$, giving
$\beta_1^{\rm O}=3/2$.

This exponent can also be derived on physical grounds as follows (see
also \cite{review_cardy}). At
the ordinary transition the density falls to zero not
exactly at the wall but would rather reach zero a distance
$\ell$ on the far side of the surface (if the density were appropriately
continued). Hence, the density on the
boundary can be computed from $\ell n_s'$. Since $\ell$ is a
microscopic distance which remains finite even at the bulk critical
point, one can compute the scaling of the surface density simply from
$n'_s=dn/dx_{\perp}|_{x_{\perp}=0}$. Thus, from 
dimensional analysis, we see that $\beta_{1}^{\rm O}=3/2$.

$\bullet$ Special Transition. In this case if $\Delta_s=0$ we see
from (\ref{int2a}) that $n_s$ scales in the same way as the bulk
density $n$, i.e. $\beta_1^{\rm Sp}=1$ \cite{mistake}. Furthermore, a
simple rewriting of (\ref{int2a}) and (\ref{int2b}) reveals the
scaling $\Delta_s \sim \Delta^{1/2}$, fixing the crossover exponent
as $\phi_1=1/2$.

$\bullet$ Surface Transition. For this case $\Delta_s<0$ and $\Delta>0$,
and hence from Eq.~(\ref{int2b}) we find
$n_s=(3/2D\lambda)[\Delta_s^2-D\Delta]$ for $0<D\Delta<\Delta_s^2$,
and $n_s=0$ for $D\Delta>\Delta_s^2$.
Hence the line in parameter space where the mean field
surface transition occurs is given by $\Delta_{s}^2=D\Delta_{\rm
critical}$, and we
then have $\beta_1^{\rm S}=1$. Note that this is the same
value as in the bulk, a standard feature of the surface transition
which is believed always to be in the same universality class as a
$d-1$ dimensional bulk transition. As we are dealing with mean field
theory this will of course yield the same exponent for the surface
transition as in the bulk. 

$\bullet$ Extraordinary Transition. 
In this case the surface density is of course non-zero both above and
below the transition. However, if we expand $n_s$ in powers of $\Delta$
for $\Delta_s<0$ and $\Delta\to 0^+$ or $0^-$, we see that these
two expansions differ at third order \cite{Eexpan},
i.e. $n_s$ has a discontinuity in
its third derivative at $\Delta=0$. Hence we identify $\beta_1^{\rm
E}=3$. To the best of our knowledge, this transition does not seem to
have been previously discussed in the literature.

However, as is the case in equilibrium critical phenomena, we expect
the extraordinary transition to be more general than the scenario
described above. In fact, the extraordinary transition is associated
with the onset of order in the bulk regardless of how the surface is
ordered. In particular, for arbitrary values of $\Delta_s$, the
surface can be ordered by applying the equivalent of a 
surface external magnetic field. For the BARW process this is simply
the surface spontaneous particle creation reaction $\emptyset\to A$.  
Extending our previous mean field analysis to cover this case
(sometimes called the normal transition), we recover precisely the
same results as obtained above, with $\beta_1^{\rm E}=3$. Hence the
important point for the
extraordinary transition (as described in
\cite{bray+moore}) is that the surface must be active at $\Delta=0$ ---
the means by which this is achieved is unimportant. 

Next, we consider the case where the bulk is {\it
exactly} critical, i.e. $\Delta$ is exactly zero, and therefore the
correlation lengths $\xi_{\parallel}$ and $\xi_{\perp}$ diverge.
In that the case
the density in the bulk decays away as $dn/dt=-\lambda n^2
\Rightarrow n \sim t^{-1}$. Hence, for the surface, we need now to
include time dependence in our analysis, and therefore
(\ref{dpmftred}) is replaced by
\begin{equation}
\label{dpmftredcrit}
\dot n=D n'' -\lambda n^2 ,
\end{equation}
where $\dot n=\partial_t n$ and the boundary condition remains
as given in Eq.~(\ref{eq:BC}).
Multiplying both sides of (\ref{dpmftredcrit}) by
$n'$ and integrating, we obtain
\begin{equation}
\label{intcrit}
\int_0^{\infty} dx_{\perp}~\dot n n' = -{1\over 2D} \Delta_s^2 n_s^2
-{1\over 3}\lambda(n^3-n_s^3) ,
\end{equation}
where we have have used the conditions $n'=0$ (in the bulk) and
the boundary condition~(\ref{eq:BC}).

$\bullet$ Exactly at the Extraordinary Transition. Here, the
density close to the wall will be in an active steady state, and hence
nearby $\dot n$ will be close to zero. However, well away from the
surface we expect to recover bulk behavior where $n'\approx 0$ and
$n\sim t^{-1}$. Hence,
to leading order the integral on the LHS of Eq.~(\ref{intcrit}) will be
zero. Therefore, from (\ref{intcrit}), we find a steady state on the
surface with $n_s\approx 3\Delta_s^2/2D\lambda$. Furthermore, we expect
that this active region will extend into the bulk, with the density
decaying away asymptotically as $x_{\perp}^{-2}$. However, assuming
the system is started with initial conditions at $t=0$ of constant
density everywhere, then after a time $t$ this region will only
extend into the bulk as far
as $x_{\perp}\sim t^{1/2}$, where we will
find a crossover to the bulk $t^{-1}$ density decay.

$\bullet$ Exactly at the Special Transition. Here, where $\Delta_s=0$, 
we see that the mean
field equation (\ref{intcrit}) is solved by $n_s=n\sim
t^{-1}$. Hence the surface density scales in the same way as in the
bulk (see also \cite{magnus}).

$\bullet$ Exactly at the Ordinary Transition. Once again if we start
with initial conditions of uniform density at $t=0$, then at later
times a depletion zone will be formed close to the surface. This zone
will again extend a distance of order $t^{1/2}$ into the bulk. The
surface scaling can now most simply be derived via 
dimensional analysis of the surface operator $\partial_{x_{\perp}}
n|_{x_{\perp}=0}$, yielding $n_s\sim t^{-3/2}$. 
   
\subsection{Beyond Mean Field Theory}
\label{bmft}

We expect that the phase diagram shown in Figure~\ref{dpps} is
generally valid 
for surface DP close to the upper critical dimension $d_c=4$. However,
in $1+1$ dimensions, where the surface is just a zero
dimensional point, the phase diagram may look rather different. For
example, for an inactive bulk, net particle production
is only possible at one point. Furthermore, since
particles will be constantly lost into the bulk,
where they will decay away exponentially quickly, it will probably
not be possible to form an active surface state (at least for finite
particle production rates). If this is the case only the ordinary
transition will be accessible in $d=1$. 
Furthermore, for arbitrary dimension, we note that a system which is
simply cut off at $x_{\perp}=0$ can also only undergo an ordinary
transition. This is a result of there being the same microscopic
reaction rates on the surface as in the bulk. In low dimensions it
becomes more and more difficult to
induce an active state (since the fluctuations become larger), and
hence if the bulk is adjusted to be at criticality, it follows that the
surface (considered independently) would be inactive. Therefore, for
the case of DP, one will only be able to find the ordinary transition
(as was certainly the case in the simulations of
\cite{dp-wall-edge}). 

The scaling forms for the survival probability (at the ordinary
transition) and correlation functions 
(at the special and ordinary transitions) have been discussed in
\cite{janssen-etal,dp-wall-edge}. For example, at the ordinary
transition, the survival probability for a cluster started on the wall
at $t=0$ has the scaling form \cite{dp-wall-edge} 
\begin{equation}
\label{P1(Delta)}
P_1(t,\Delta)=|\Delta|^{\beta_1^{\rm O}}\varphi_1(t/\xi_{\parallel}) .
\end{equation}
Hence the average lifetime of finite clusters at the ordinary transition,
$\langle t \rangle \sim |\Delta|^{-\tau_1^{\rm O}}$, satisfies 
$\tau_1^{\rm O} = \nu_\parallel - \beta^{\rm O}_1$, a straightforward
generalization of the bulk result. Previous series expansions
in $1+1$ dimensions have indicated a value for $\tau_1^{\rm O}$
equal to unity \cite{essam-etal:1996}, although very recent (and even
more accurate) series
results \cite{newjensen} have cast some doubt on this conclusion. No
theoretical explanation for why $\tau_1^{\rm O}$ should be equal to
unity has emerged.

The bulk density $n({\bf x},t)$ for a
cluster initiated on the wall at $t=0$ is given by
\cite{janssen-etal,dp-wall-edge}
\begin{eqnarray}
\label{jancorr}
& & n_1^{\rm O}(x,t,\Delta) = |\Delta|^{\beta_1^{\rm O} + \beta}
        f_1  \left[x/{\xi_\perp}, \, {t}/{\xi_\parallel}\right]. \\
& & n^{\rm Sp}_{1}(x,t,\Delta,\Delta_s) = |\Delta|^{\beta_1^{\rm Sp} +
        \beta} \tilde f_1  \left[x/{\xi_\perp}, \, {t}/{\xi_\parallel,
        \, \Delta_s/|\Delta|^{\phi_1} }\right] , \nonumber
\end{eqnarray}
where the surface exponents have been calculated to $O(\epsilon=4-d)$
in \cite{janssen-etal} giving $\beta_1^{\rm O}=3/2-7\epsilon/48$, 
$\beta_1^{\rm Sp}=1-\epsilon/4$, and $\phi_1=1/2-\epsilon/16$.
The first of the expressions in (\ref{jancorr}) refers to the ordinary
transition and the second to the special transition. Crudely speaking,
the $\Delta$-prefactor in (\ref{jancorr}) comes from  
(\ref{P1(Delta)}) for the probability that an infinite cluster can be
grown from the seed, and from (\ref{n(Delta)}) for the conditional
probability that the point (${\bf x}$, $t$) belongs to this
cluster. At the ordinary transition, for example, it is then
straightforward to 
derive hyperscaling relations for the mass of finite clusters which
are seeded on the wall. This mass scales as 
$\langle s_1 \rangle \sim |\Delta|^{-\gamma_1^{\rm O}}$, and as shown
in \cite{dp-wall-edge} one finds
\begin{equation}
        \nu_{\parallel}+d\nu_{\perp}=
        \beta_{1}^{\rm O}+\beta+\gamma^{\rm O}_{1} .
        \label{surf_hyperscalingDP}
\end{equation}
For a more detailed discussion of the DP ordinary transition, including
other types of hyperscaling, we refer to Ref.~\cite{dp-wall-edge}. 

\section{Surface BARW}
\label{BARWsurface}

We now turn our attention to the main object of this paper, an
understanding of the surface critical properties of BARW. We will
begin by discussing the BARW surface phase diagram in various
dimensions. The basic idea is that on the surface we may include not
only the usual branching and annihilation reactions but potentially
also a parity symmetry breaking $A\to\emptyset$ reaction. Depending on 
whether or not the $A\to\emptyset$ reaction is actually present, we
may then expect different surface universality classes according to
whether the symmetry of the bulk is broken or respected at the
surface. A similar situation in an equilibrium system has recently
been analyzed in Ref.~\cite{drewitz}. We will find that the
competition between the parity breaking $A\to\emptyset$ reaction and
the BARW processes gives rise to some interesting phase diagrams.

\subsection{Mean field phase diagram}
\label{mfbarw}

\begin{figure}
\begin{center}
\leavevmode
\vbox{
\epsfxsize=2in
\epsffile{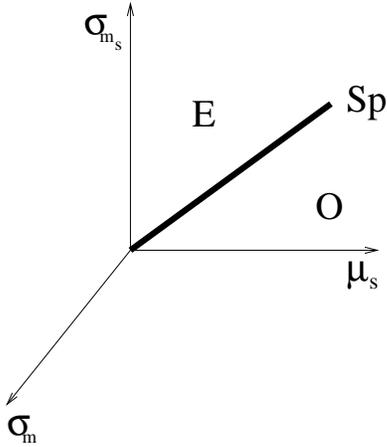}}
\end{center}
\caption{Schematic mean field surface phase diagram for BARW. See text
for an explanation of the labeling.} 
\label{psurf}
\end{figure}

The surface phase diagram for the mean field theory of BARW (valid for
$d>d_c=2$) is shown in Figure~\ref{psurf}. Here $\sigma_m$,
$\sigma_{m_s}$ are the rates for the branching processes $A\to (m+1)A$
in the bulk and at the
surface, respectively, and $\mu_s$ is the rate for the surface
spontaneous annihilation reaction $A\to\emptyset$. Otherwise, the
labeling is the same as that for the DP phase diagram (see
Figure~\ref{dpps}).  

The first feature to note is that the bulk is either active
($\sigma_m>0$) or critical ($\sigma_m=0$), but never inactive. Hence, 
unlike DP, there is no possibility of finding a surface transition,
where the surface is critical with the bulk inactive. For the case
where $\sigma_m=\mu_s=0$, we expect that for any finite value of the
surface branching, the surface will become active. This corresponds to
the extraordinary transition with an active surface and critical
bulk. On the other hand for $\sigma_m=\sigma_{m_s}=0$ and $\mu_s>0$, the
density at an (isolated) surface would decay away exponentially
quickly due to the
$A\to\emptyset$ reaction. Hence the bulk is critical, with the surface
inactive; i.e. the ordinary transition. Consequently with $\sigma_m=0$,
but both $\mu_s$ and $\sigma_{m_s}$ non-zero, there should be a line of
special transitions dividing the extraordinary and ordinary
regions. This explains the general features of the phase diagram in
Figure~\ref{psurf}. 

At a more quantitative level, the mean field equation for BARW is very
similar to that for DP:
\begin{equation}
	\label{barwmfteq}
	\partial_t n = D\nabla^2 n -\Delta n -\lambda n^2 ,
\end{equation}
with the boundary condition 
\begin{equation}
	D \partial_{x_{\perp}}n|_{x_{\perp}=0}= \Delta_s
	n|_{x_{\perp}=0}. \label{eq:BCbarwgeneral}
\end{equation}
However the values of the $\Delta$, $\Delta_s$ parameters are now
different: $\Delta=-m\sigma_m$ and $\Delta_s=-m\sigma_{m_s}+\mu_s$.
The fact that $\Delta$ is always non-positive excludes any possibility
of a surface transition. Otherwise we expect the same mean
field exponents as in DP for the special and ordinary transitions
\cite{mistake} (except for the $\beta_{1,{\rm seed}}$ exponents, see
below). However, the non-positivity of $\Delta$ also leads to an
ambiguity associated with the definition of $\beta_{1,{\rm dens}}^{\rm
E}$: we would have to know the behavior of the surface density on both
sides of the extraordinary transition if we wanted to isolate the
discontinuity and extract the exponent.

We can also consider the mean field behavior of the $\beta_{1,{\rm seed}}$
exponents, which is very different from the corresponding behavior in
DP. Consider placing two particles next to the surface at
$t=0$. From the recurrence properties of
random walks we see that, regardless of the reaction rates on the
surface or in the bulk, there is a finite chance that the two
particles will never meet again. Hence the survival probability is
{\it non-zero} and thus $\beta_{1,{\rm seed}}=0$ in mean
field theory for the ordinary and special transitions. 

\subsection{Phase diagram in $1+1$ dimensions}
\label{d=1}

\begin{figure}
\begin{center}
\leavevmode
\vbox{
\epsfxsize=3in
\epsffile{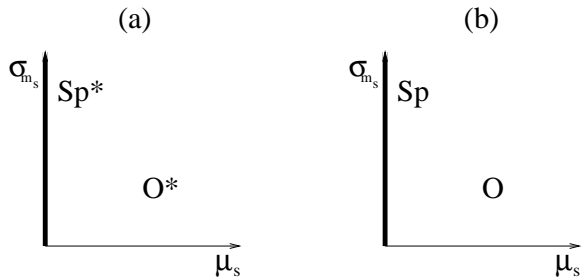}}
\end{center}
\caption{Schematic surface phase diagrams for BARW in $d=1$ for (a)
$\sigma_m<\sigma_{m,{\rm critical}}$, and (b) $\sigma_m=\sigma_{m,{\rm
critical}}$. See text for an explanation of the labeling.} 
\label{psurf1d}
\end{figure}

Next, we turn our attention to the phase diagram for $1+1$ dimensions
shown in Figure~\ref{psurf1d}. Although we will make a few remarks
below, we will
postpone a proper justification until we have discussed the
appropriate field theory in 
Section~\ref{fieldth}. The phase diagram looks quite different from
its mean field analog due in part to the shift of the bulk critical
point away from zero branching rate, but also due to the absence of
any extraordinary transition (for finite reaction rates). Physically,
this is due to the fact that
excess particle production (with a finite reaction rate) at a zero
dimensional surface is simply not
efficient enough to generate an active state, due to leakage into the
critical bulk (which for $\sigma_m<\sigma_{m,{\rm critical}}$ is
controlled by the fixed point of the $A+A\to\emptyset$ reaction
\cite{cardy-tauber}). 

However, for infinite
branching rates and/or if the reaction $\emptyset\to A$ is added at
the surface, then an extraordinary transition should become
accessible, although we will not consider this case in any further
detail. 
The other main features of the phase diagram in $1+1$ dimensions 
are outlined below. 

$\bullet$ Sp*: For
$\sigma_m<\sigma_{m,{\rm critical}}$ and $\mu_s=0$, the system is
controlled 
everywhere by the annihilation fixed point. In that case one has the
special transition, but now in a slightly different sense to what we
have seen before. In this region it is {\it not} possible to obtain an
active state either on the surface or in the bulk, by small
changes in the bulk and/or surface branching rates. Hence this
``transition'' is actually 
entirely controlled by the $A+A\to\emptyset$ process with the
branching playing essentially no role. Thus we have marked this
``transition'' as Sp* in Figure~\ref{psurf1d}a. This simpler and
analytically 
tractable case has already been extensively analyzed in
Ref.~\cite{magnus}. We will postpone further theoretical
discussion until Section~\ref{aares}.

$\bullet$ Sp: Next we consider the special transition at
$\sigma_m=\sigma_{m,{\rm critical}}$, $\mu_s=0$. This transition
borders the bulk active phase, and hence will belong to a
quite different universality class to that described immediately
above, and will instead be similar to the special transitions
discussed in earlier sections (although fluctuations will now be
very important for this $1+1$ dimensional case). 

$\bullet$ O*: For $\mu_s>0$ and $\sigma_m<\sigma_{m,{\rm critical}}$,
the presence of the $A\to\emptyset$ reaction on the surface gives rise
to an ordinary ``transition''. However, as explained above, the 
branching process again plays essentially no role here. Further
details of this O* ``transition'' are provided in Section~\ref{aares}.

$\bullet$ O: Finally, at
$\sigma_m=\sigma_{m,{\rm critical}}$, $\mu_s>0$, we expect an
ordinary transition similar in character to
the ordinary transitions discussed in previous sections (although in
this $1+1$ dimensional case the fluctuations are again very
important). 

\subsection{Scaling Theory}
\label{corrbarw}

In this section we construct a scaling theory for the survival
probabilities and
correlation functions at the Sp and O transitions. This scaling
theory is certainly valid for the mean field regime but there are,
however, subtleties involved in its application to the fluctuation
dominated regime for
$d\leq 2$. The extent of its validity in that region will be discussed
in detail in the next section. When writing down this scaling 
theory we must also bear in mind the
important distinction between the $\beta_{1,{\rm dens}}$ and
$\beta_{1,{\rm seed}}$ exponents. We begin by giving
a scaling form for the survival probability $P_1(t,\Delta)$, where
$\Delta$ is the deviation from bulk BARW criticality. For
example, at the ordinary transition, for a seed placed on the wall at
${\bf x}={\bf 0}$, $t=0$, we have 
\begin{equation}
\label{barwsurvwall}
P_1(t,\Delta)=|\Delta|^{\beta^{\rm O}_{1,{\rm seed}}}
\Phi_1(t/\xi_{\parallel}). 
\end{equation}
It is then straightforward to compute the average lifetime of finite
clusters, $\langle t \rangle \sim |\Delta|^{-\tau_1}$, where
$\tau_1^{\rm O} = \nu_\parallel - \beta_{1, \rm seed}^{\rm O}$,
just as in the case of DP.  

Next, we consider the coarse-grained particle density 
$n_1$ at the point (${\bf x}$, $t$) for a cluster grown
from a seed located next to the wall at ${\bf x}={\bf 0}$, $t=0$. At
the ordinary transition we have
\begin{equation}
        \label{ansatz_rho_wall}
        n_{1}(x,t,\Delta) = |\Delta|^{\beta^{\rm O}_{1, \rm seed} +
        \beta_{\rm dens}}
        g_1  \left(x/{\xi_\perp}, \, {t}/{\xi_\parallel}\right) .
\end{equation}
As was the case for DP, the $\Delta$-prefactor in
(\ref{ansatz_rho_wall}) comes from  
(\ref{barwsurvwall}) for the probability that an infinite cluster can be
grown from the seed, and from (\ref{n(Delta)}) for the (conditional)
probability that the point (${\bf x}$, $t$) belongs to this cluster.
The shape of the cluster is governed by the scaling function $g_1$ and
we assume that the density is measured at a finite angle away from the
wall. If the density is measured along the wall, we have instead
\begin{equation}
        \label{ansatz_rho_wall_wall}
        n_{11}(x,t,\Delta) =
        |\Delta|^{\beta^{\rm O}_{1, \rm seed} + \beta^{\rm O}_{1, \rm dens}}
        g_{11} \left(x/{\xi_\perp}, \,
        t/{\xi_\parallel}\right) ,
\end{equation}
as we pick up a factor $|\Delta|^{\beta^{\rm O}_{1, \rm dens}}$ rather
than $|\Delta|^{\beta_{\rm dens}}$ from the probability that (${\bf x}$,
$t$) at the wall belongs to the cluster. 

The above correlation functions need only be modified slightly to be
valid at the special transition. If $\Delta_s$ (the deviation
of the surface from criticality) is a relevant parameter, then we must
take care to
include the extra variable $\Delta_s/|\Delta|^{\phi_1}$ in the scaling
function. The scaling form replacing (\ref{ansatz_rho_wall}) then
becomes 
\begin{equation}
        \label{ansatz_rho_wall_sp}
        n_{1}(x,t,\Delta) = |\Delta|^{\beta^{\rm Sp}_{1, \rm seed} +
        \beta_{\rm dens}}
        \tilde g_1  \left[x/{\xi_\perp}, \, {t}/{\xi_\parallel}, \,
        \Delta_s/|\Delta|^{\phi_1} \right] ,
\end{equation}
where $\phi_1$ is a crossover exponent associated with the
multicritical special transition. Similarly 
Eq.~(\ref{ansatz_rho_wall_wall}) is replaced by
\begin{equation}
        \label{ansatz_rho_wall_wall_sp}
        n_{11}(x,t,\Delta) =
        |\Delta|^{\beta^{\rm Sp}_{1, \rm seed} + \beta^{\rm Sp}_{1,
        \rm dens}} 
        \tilde g_{11}\left[x/{\xi_\perp}, 
        t/{\xi_\parallel},
        \Delta_s/|\Delta|^{\phi_1}\right]\hspace{-0.5mm}. 
\end{equation}
Note, however, that there are subtleties concerning the special
transition in $1+1$ dimensions which will be discussed in
Section~\ref{ibcfth}. 

At the ordinary transition, for example, we can use the above scaling
forms to derive some further exponent equalities. The average 
size of finite clusters 
\begin{equation}
        \label{size_wall}
        \langle s_1 \rangle \sim |\Delta|^{-\gamma^{\rm O}_1} ,
\end{equation}
follows from integrating the cluster density 
(\ref{ansatz_rho_wall}) over space and time,
where the surface (susceptibility) exponent $\gamma^{\rm O}_1$ is
related to the previously defined exponents via 
\begin{equation}
        \nu_{\parallel}+d\nu_{\perp}=
        \beta^{\rm O}_{1, \rm seed}+\beta_{\rm dens}+\gamma^{\rm O}_{1} .
        \label{surf_hyperscaling}
\end{equation}
Analogously, by integrating the cluster wall density 
(\ref{ansatz_rho_wall_wall}) over the ($d-1$)-dimensional wall and 
time, we obtain the average size of finite clusters on the wall
\begin{equation}
        \label{size_wall_wall}
        \langle s_{1,1} \rangle \sim |\Delta|^{-\gamma^{\rm O}_{1,1}} ,
\end{equation}
where
\begin{equation}
        \nu_{\parallel}+(d-1)\nu_{\perp}=
          \beta^{\rm O}_{1, \rm seed}+\beta^{\rm O}_{1, \rm dens}+
          \gamma^{\rm O}_{1,1}.
        \label{surf_hyperscaling_wall_wall}
\end{equation}
Note that if the $\gamma$ susceptibility exponents obtained from 
(\ref{surf_hyperscaling}) and (\ref{surf_hyperscaling_wall_wall})
are negative, then they should be replaced by zero in (\ref{size_wall}) 
and (\ref{size_wall_wall}). 

\subsection{Field Theory}
\label{fieldth}

In order to properly understand the effects of
fluctuations, and to justify some of the scaling forms proposed in
the last section, we now turn to the development of a field theory for 
surface BARW.  
We will begin by reviewing the field theory for BARW in the 
bulk, before moving on to derive the appropriate surface actions.
The bulk BARW action, written in terms of the response field
$\hat\psi({\bf x},t)$ and the  
``density'' field $\psi({\bf x},t)$, is given by \cite{cardy-tauber}
\begin{eqnarray}
& & S^{\rm bare}_{\rm bulk}[\psi,\hat\psi;\tau]=
\int d^dx \left[ \int_0^{\tau} dt \left[\hat\psi({\bf
x},t)[\partial_t- 
D{\bf\nabla}^2]\psi({\bf x},t) \right.\right. \nonumber \\
& & \left.\left.\qquad\qquad\qquad\qquad\qquad\quad
-\lambda[1-\hat\psi({\bf x},t)^2]\psi({\bf x},t)^2
\right. \right. \label{baweaction} \\
& & \left.\left.\qquad\qquad\qquad\qquad\quad + \sigma_m[1-\hat\psi({\bf
x},t)^m]\hat\psi({\bf 
x},t)\psi({\bf x},t)\right] \right. \nonumber \\
& & \qquad\qquad\qquad\qquad\qquad\quad\left.
-\psi({\bf x},\tau)-n_0\hat\psi({\bf x},0)\right] \nonumber .
\end{eqnarray}
Here the terms on the first line of (\ref{baweaction}) represent
diffusion of the particles (with continuum diffusion constant
$D$). The second line describes the annihilation reaction (with
continuum rate $\lambda$), whilst the terms on the third line
represent the branching process (with continuum rate $\sigma_m$). The
final two terms represent, respectively,  a contribution due to the
projection state (see \cite{peliti}), and the initial condition (an
uncorrelated Poisson distribution with mean $n_0$). In the
following we will restrict ourselves to the case of {\it even} $m$,
since it is known that the odd $m$ case belongs to the DP universality
class \cite{cardy-tauber}.

The action given in (\ref{baweaction}) is a {\it bare}
action. In order to properly include fluctuation effects one must be 
careful to include processes generated by a combination of branching
and annihilation. In other words in addition to the process $A\to
(m+1)A$, the reactions $A\to (m-1)A$, \ldots, $A\to 3A$ need to be
included. These considerations lead to the full action
\begin{eqnarray}
& & S_{\rm bulk}[\psi,\hat\psi;\tau]=
\int d^dx \left[ \int_0^{\tau} dt \left[\hat\psi({\bf x},t)
[\partial_t- D{\bf\nabla}^2]\psi({\bf x},t) \right.\right. \nonumber \\
& & \left.\left.\qquad\qquad\qquad
+ \sum_{l=1}^{m/2}\sigma_{2l}[1-\hat\psi({\bf x},t)^{2l}]
\hat\psi({\bf x},t)\psi({\bf x},t) \right. \right. \label{barwbulk} \\
& & \left.\left. \qquad\quad -\lambda[1-\hat\psi({\bf
x},t)^2]\psi({\bf x},t)^2]
-\psi({\bf x},\tau)-n_0\hat\psi({\bf x},0)] \right.\right. \nonumber .
\end{eqnarray}
Notice also that (for {\it even $m$}) the action (\ref{barwbulk})
is invariant under the ``parity'' transformation
\begin{equation}
\label{parity}
\hat\psi({\bf x},t)\to -\hat\psi({\bf x},t), \qquad \psi({\bf x},t)\to
-\psi({\bf x},t) .
\end{equation}
This symmetry corresponds physically to particle conservation modulo
$2$. The presence of this extra symmetry now takes the system away
from the DP universality class, and into a new class: that of
branching-annihilating random walks with an even number of offspring. 

Close to the upper critical dimension $d_c=2$, the renormalization of
the above action is quite straightforward (here we quote the results from
\cite{cardy-tauber}). Only the branching and annihilation rates need be 
renormalized, and in particular there are no diffusion constant or field
renormalizations. Furthermore, if we are close to the annihilation fixed
point, then the RG eigenvalue of the branching parameter becomes
$y_{\sigma_m}=2-m(m+1)\epsilon/2+O(\epsilon^2)$, where $\epsilon
=2-d$. Hence we see that the {\it lowest} branching process is
actually the most relevant. Therefore, close to $2$ dimensions where
the branching remains relevant, we expect to find an {\it active}
state even for very 
small values of the branching (in agreement with the mean field phase
diagram). Furthermore, the fact that there is only one eigenvalue to
compute perturbatively (as the renormalization of the annihilation rate
can be performed to all orders \cite{lee}), means that there is only one
independent exponent. Hence, close to $2$ dimensions, the order
parameter exponent can be shown to be $\beta_{\rm dens}=d\nu_{\perp}$.

However, inspection of the above RG eigenvalue $y_{\sigma_m}$ shows
that it eventually becomes negative (if the one loop result is to be 
believed). In that case we expect a major change in the behavior of
the system, since the branching process will no longer be relevant at
the annihilation fixed point. The critical transition point is then
shifted with the active state only being present for values of the
branching greater than some positive critical value (as indicated in
Figure~\ref{pbulk}b).   
Consequently, we see that there is a second critical
dimension $d_c'\approx 4/3$ whose presence immediately rules out any
possibility of accessing the non-trivial behavior expected in $d=1$
via perturbative epsilon expansions down from $d=2$. Instead cruder
techniques (such as the loop expansion in fixed dimension) must be
employed \cite{cardy-tauber}. 

We now turn to the derivation of the surface actions appropriate
for the cases $\mu_s=0$ and $\mu_s\neq 0$.

\subsubsection{$\mu_s=0$ Field Theory}
\label{ibcfth}

Starting from an appropriate master equation for the system on a lattice, 
the form of the surface action can be derived using standard techniques 
\cite{peliti,lee}. After mapping to the continuum theory,
we find the bare action $S^{\rm bare}=S^{\rm bare}_{\rm bulk}+S^{\rm
bare}_1$, with $S_{\rm bulk}$ given by (\ref{baweaction}) and
\begin{equation}
\label{IBCaction}
S_1^{\rm bare}= \int d^{d-1}x_{\parallel} \int_0^{\tau} dt ~ \left[
\sigma_{m_s} [1-\hat\psi^m_s] \hat\psi_s\psi_s\right] ,
\end{equation}
where $\hat\psi_s=\hat\psi({\bf x_{\parallel}},x_{\perp}=0,t)$ and
$\psi_s=\psi({\bf x_{\parallel}},x_{\perp}=0,t)$. Note that the terms
representing the annihilation reaction $A+A\to\emptyset$ are {\it
irrelevant} on the surface close to the upper critical dimension..
The classical field equations for the above action can be derived 
by taking the variational derivatives of the action $S^{\rm bare}
=S^{\rm bare}_{\rm bulk}+S^{\rm bare}_1$ with respect to the fields
$\hat\psi$ and $\psi$. These 
equations are solved by $\hat\psi=\hat\psi_s=1$, with $\psi$
satisfying
\begin{equation}
\partial_t\psi=D\nabla^2\psi-\Delta\psi-2\lambda\psi^2 ,
\end{equation}
where $\Delta=-m\sigma_m$, and with the boundary
condition $D\partial_{x_{\perp}}\psi|_{x_{\perp}=0}=\Delta_s\psi_s$,
where $\Delta_s=-m\sigma_{m_s}$. These mean field results are in
agreement with our analysis in Section~\ref{mfbarw}.
Furthermore, we note that a boundary term of
the form $\hat\psi_s \partial_{x_{\perp}}\psi|_{x_{\perp}=0}$,
although marginal 
from power counting arguments, is actually always redundant (even in
the regime where mean field theory no longer applies). This is also the
case for the surface action in DP (see \cite{janssen-etal}).  

However, if we are properly to include fluctuation effects, we
must again take
care to include surface terms generated by a combination of branching
and annihilation (as in the bulk). This leads to the full surface
action 
\begin{equation}
\label{IBCactionfull}
S_1= \int d^{d-1}x_{\parallel} \int_0^{\tau} dt ~ \left[
\sum_{l=1}^{m/2}\sigma_{2l_{s}}[1-\hat\psi_s^{2l}]
\hat\psi_s\psi_s \right] .
\end{equation}
Note also that the parity symmetry (\ref{parity}) is preserved
for the $\mu_s=0$ model at the wall, as well as in the bulk.

Power counting on the above action reveals that the
surface branching rates $\sigma_{2l_s}$ all have naive dimension
$[\sigma_{2l_s}]\sim \kappa^1$, where $\kappa$ denotes an inverse
length scale. However, below $2$ dimensions this scaling dimension
will be renormalized downwards (this can be seen physically as a
result of processes like $A\to 3A\to A$ rendering the
branching process less efficient). As a result of this
renormalization, we expect the lowest generated process
(i.e. with $l=1$ in Eq.~(\ref{IBCactionfull})) will become the {\it
most} relevant (as it was in the bulk). Nevertheless, despite this
downward renormalization, close enough to $2$ dimensions,  
the scaling dimension of the most relevant coupling $\sigma_{2_s}$
will remain positive, and thus under the
RG will flow to $\infty$ for all non-zero starting values. 
This state of affairs corresponds to the extraordinary transition,
where the surface is {\it active} while the bulk is critical.
On the other hand, at bulk criticality and with $\sigma_{2_s}=0$, we
have a multicritical special transition point.
In this case, after writing down and solving the
appropriate RG equations (exactly along the lines of
\cite{cardy-tauber,janssen-etal}), one can derive the scaling results
for the density quoted in Section~\ref{corrbarw}, where we can see
that the independent renormalization of $\sigma_{2_s}$ contributes to
the crossover exponent $\phi_1$. Furthermore, since there is
no field renormalization (either at the surface or in the bulk), this
implies that the exponent $\beta_{1,{\rm dens}}^{\rm Sp}$ is just the
same as in the bulk, i.e. $\beta_{1,{\rm dens}}^{\rm Sp}=\beta_{\rm
dens}$. However, we must again stress that this result is only true
close to $d=2$. 

The situation in $d=1$ is rather different, partly due to the
shift of the bulk critical point away from $\sigma_m=0$. This
means that the $d=1$ transition at $\sigma_m=\sigma_{m,{\rm
critical}}$ {\it cannot} be based on perturbative epsilon expansion
calculations down 
from $2$ dimensions. However, we can say a little more
if we first consider the regime $\sigma_m<\sigma_{m,{\rm 
critical}}$ in $d=1$, where the bulk is controlled solely by the
$A+A\to\emptyset$ reaction. In that case we expect the scaling
dimension of all the $\sigma_{2l_s}$ to be negative in $d=1$,
following the downwards trend in the renormalization mentioned
above. In that case surface branching is then {\it irrelevant} in
$d=1$ leading to the Sp* special ``transition''. Similarly, at the Sp
transition at $\sigma_m=\sigma_{m,{\rm critical}}$, we might again 
expect the $\sigma_{2l_s}$ to be irrelevant. 
This will be reflected in the scaling functions for the density, where
(unusually) the crossover term $\Delta_s/|\Delta|^{\phi_1}$ will now
be absent. 
However, the surface exponents here will presumably be unrelated to
the bulk exponents, since the absence of field renormalization
mentioned above is not expected to hold all the way down to
$d=1$. 

Hence, if the above scenario is correct, we do not expect to
see an extraordinary transition in $d=1$ for any finite value of the
surface 
branching, since the surface branching will always be irrelevant. We
have confirmed this analysis 
numerically: our simulations have found no evidence of an active
surface state for $\sigma_m\leq\sigma_{m,{\rm critical}}$ even for
very high values of the surface branching
parameter in a fermionic lattice model in $1+1$
dimensions (see Section~\ref{numres} for further details). 

\subsubsection{$\mu_s\neq 0$ Field Theory}
\label{rbcfth}

In this case the reaction $A\to\emptyset$ {\it is} now possible, but
only at sites on the wall. In the bosonic field theory language
employed above, we have the bare action $S^{\rm bare}=S_{\rm
bulk}^{\rm bare}+S^{\rm bare}_2$, where 
\begin{eqnarray}
\label{RBCaction}
& & S^{\rm bare}_2= \int d^{d-1}x_{\parallel} \int_0^{\tau} dt ~ [
\sigma_{m_s} [1-\hat\psi^m_s] \hat\psi_s\psi_s \\ \nonumber & &
\qquad\qquad\qquad\qquad\qquad\qquad + \mu_s[\hat\psi_s-1]\psi_s] .
\end{eqnarray}
The symmetry (\ref{parity}) is now broken by the surface
term proportional to $\mu_s$, which describes the $A\to\emptyset$
reaction.
Repeating our derivation of the classical (mean field) equation for
the $\psi$ density field, we find
\begin{equation}
\partial_t\psi=D\nabla^2\psi-\Delta\psi-2\lambda\psi^2 ,
\end{equation}
where $\Delta=-m\sigma_m$, and with the boundary condition
$D\partial_{x_{\perp}}\psi|_{x_{\perp}=0}=\Delta_s\psi_s$, where
$\Delta_s=-m\sigma_{m_s}+\mu_s$. This is in agreement with the mean
field analysis given in Section~\ref{mfbarw}.
Note that a boundary term of the form
$\hat\psi_s\partial_{x_{\perp}}\psi|_{x_{\perp}=0}$ is again always
redundant.

The action (\ref{RBCaction}) is a bare action whose terms simply
represent the reactions $A\to (m+1)A$ and $A\to\emptyset$ at the
surface. Clearly, however, from these two reactions we can generate
the hierarchy of processes $A\to mA$, $A\to (m-1)A$, \ldots, $A\to
2A$. Hence we must replace the above bare surface action with
\begin{eqnarray}
\label{RBCgenaction}
& & S_2= \int d^{d-1}x_{\parallel} \int_0^{\tau} dt ~ \left[
\sum_{l=1}^m  ~ \sigma_{l_s} [1-\hat\psi^l_s]
\hat\psi_s\psi_s\right. \\ \nonumber & & \left.
\qquad\qquad\qquad\qquad\qquad\qquad + \mu_s[\hat\psi_s-1]\psi_s\right] .
\end{eqnarray}

The renormalization of the action (\ref{RBCgenaction}) is now
somewhat different from  the $\mu_s=0$ case. We again expect that
we need only keep the lowest generated branching term on the surface,
namely that with $l=1$ in (\ref{RBCgenaction}). 
As before, we expect fluctuations to lower the
scaling dimension of this coupling from its mean field
value (although actually in $d=2$ this suppression will only be
logarithmic).  On the other hand, the efficacy of the $A\to\emptyset$
reaction is certainly {\it not} reduced by fluctuations. Hence, we
expect that $\Delta_s\to \mu_s-\sigma_{1_s}$ will always run to the
fixed point at $\infty$, corresponding to the ordinary transition. In
that case the surface $\beta_{1,{\rm dens}}^{\rm O}$ exponent
is again simply related to the bulk result due to the absence of any
surface field renormalization. This exponent can be
computed from the scaling of the surface operator $\partial
\psi/\partial x_{\perp}|_{x_{\perp}=0}$, where the $x_{\perp}$
derivative simply brings out an
extra factor of $\nu_{\perp}$ from the scaling function, giving 
$\beta_{1,{\rm dens}}^{\rm O}=\beta_{\rm
dens}+\nu_{\perp}$. Using the result
$\beta_{\rm dens}=d\nu_{\perp}$ from \cite{cardy-tauber}, we see that
$\beta_{1,{\rm dens}}^{\rm O}=(d+1)\nu_{\perp}$. Again we stress that
this result is only true close to $2$ dimensions.
The more interesting transition at
$\sigma_m=\sigma_{m,{\rm critical}}$ in $d=1$ is not perturbatively
accessible in epsilon expansions down from $d=2$. Nevertheless, we
still expect 
the same general picture to hold with the surface branching always
being irrelevant, leading to the O* ($\sigma_m<\sigma_{m,{\rm
critical}})$ or O ($\sigma_m=\sigma_{m,{\rm critical}})$
transitions. 

\subsubsection{Discussion}
\label{fthdisc}

{}From the above analysis we can understand the structure of the
phase 
diagram close to $2$ dimensions. For $\mu_s=0$ this is similar to the
mean field picture, with a special transition point at
$\sigma_m=\sigma_{m_s}=0$, and with the extraordinary transition for
$\sigma_m=0$, $\sigma_{m_s}>0$. On the other hand, for $\mu_s>0$, the
picture is very different 
from mean field theory, with renormalization effects ensuring
that only the ordinary transition is accessible. However, actually in
$d=2$ this might be hard to observe, since in that case the
surface branching is only marginally less relevant.

One would now like to use the actions (\ref{IBCaction}) and
(\ref{RBCaction}) as the starting point for a field theoretic 
investigation of the $\sigma_m=\sigma_{m,{\rm critical}}$ transitions
in $d=1$, 
where one would like to identify {\it two independent, non-trivial}
surface $\beta_1$ exponents (a feature which is certainly
indicated by our simulations, see Section~\ref{numres}). 
Surprisingly, our numerical results also indicate that these surface
$\beta_1$ exponents
``swap'' if the $\mu_s=0$ and $\mu_s\neq 0$ cases are interchanged
(i.e. $\beta_{1,{\rm dens}}^{\rm O}=\beta_{1,{\rm seed}}^{\rm Sp}$ and
$\beta_{1,{\rm seed}}^{\rm O}=\beta_{1,{\rm dens}}^{\rm Sp}$).
These interesting results certainly merit further analysis. 
Unfortunately the use of field theoretic techniques here will
be plagued by precisely the same
problems as afflicted the bulk calculation, namely the appearance of a
second critical dimension $d_c'$. Hence one would be forced into
using uncontrolled techniques (such as the truncated loop expansion in
fixed 
dimension) whose values for the bulk exponents are known to be 
in rather poor agreement with numerics \cite{cardy-tauber}. 
Furthermore, field and diffusion constant renormalizations, which will
be of considerable importance in $d=1$, are not adequately taken into
account in the truncated loop theory. In fact these renormalizations
only appear at two loop order. Unfortunately the authors of
\cite{cardy-tauber} were unable to show that a meaningful
truncated loop theory exists at all at the level of two loops.
In addition, further technical difficulties exist for $d<d_c'$
involving dangerous irrelevant variables, which have so far
prevented a derivation of scaling relations {\it at} criticality even
in the bulk. In the light of these problems we have not attempted to
extend the truncated loop analysis to surface BARW. 

\subsubsection{Results from $A+A\to\emptyset$}
\label{aares}

The $1+1$ dimensional regime which should more amenable to field
theoretic analysis 
is when both the bulk and surface branching processes are
unimportant, and hence we should be able to use results derived solely
from the $A+A\to\emptyset$ reaction. This is the case 
for the region $\sigma_m<\sigma_{m,{\rm critical}}$ in $d=1$
(see Figure~\ref{psurf1d}a). The Sp* ``transition'' has been fully
analyzed in Ref.~\cite{magnus}, which predicts
a $t^{-1/2}$ decay both in the bulk and at the wall, with a
density excess at the wall. We note that the critical Sp*
state in $d=1$ can be characterized in two ways: as a decaying
density, or as a 
survival probability. If we place two particles close together in the
bulk at $t=0$, then, from simple random walk 
theory, the probability these particles are still alive at time $t$
scales as $t^{-1/2}$. Hence, in the bulk, these two ways of
characterizing this phase scale in the same way. However, if the two
particles are released next to the wall, then it is easy to show
(using the method of images) that the survival probability now decays
as $t^{-1}$ \cite{benav,redner}. Therefore, these two
characterizations of the Sp* state do {\it not} scale in the same way
close to the wall.  

Since the bulk at the $d=1$ O* transition is controlled solely by the
reaction $A+A\to\emptyset$ , its properties can also be inferred. The
relevant surface operator is again just
$\partial_{x_{\perp}}\psi|_{x_{\perp}=0}$. Therefore,
since distance still scales as $[x]\sim
t^{1/2}$, we can obtain the required surface scaling from simple
dimensional analysis. Since the bulk $d=1$ density decays as
$t^{-1/2}$, we see that the surface density must decay as $t^{-1}$.

However, we must emphasize again at this point that the methods and
results mentioned here are only applicable where both the surface and bulk
branching processes are {\it unimportant}.
Unfortunately, therefore, the more interesting Sp and O transitions
in $1+1$ dimensions remain out of reach. 

Hence, given the fundamental difficulties associated with the field 
theory, it seems fruitful to search for alternative approaches to the
problem which might shed some further light on the interesting
properties of the surface $\beta_1$ exponents in $d=1$.
One such alternative is provided by the theory of quantum spin
Hamiltonians, to which we turn in the next section. 

\subsection{Exact Results}
\label{exres}

In this section we will derive some exact results for the surface
$\beta_1$ exponents in $1+1$ dimensions at the O and Sp
transitions. The methods are a straightforward
extension of the work in \cite{schutz2,schutz1}. The starting point is
the following set of rules for BARW with $m=2$ in $1+1$ dimensions:
\begin{eqnarray}
& \emptyset A \leftrightarrow A \emptyset & ~~{\rm with~rate}~D/2
\nonumber \\
\label{fbarwrules}
& A A \to \emptyset\emptyset & ~~{\rm with~rate}~\lambda \\
& \emptyset A \emptyset \leftrightarrow A A A~~{\rm and}~~\emptyset A A
\leftrightarrow A A \emptyset & ~~{\rm with~rate}~\alpha/2 . \nonumber
\end{eqnarray}
Note that these rules are fermionic in character (no more than one
particle per site is permitted) in contrast to the bosonic rules
employed in the derivation of the earlier field theory.
The model described in (\ref{fbarwrules}) can be transformed into a
spin picture by writing the
configuration of a semi-infinite system as a vector
$|s_1,s_2,s_3,\ldots\rangle$, where $s_i=1/2$ if the $i$-th site is
empty, and $s_i=-1/2$ if that site is occupied. Hence the system ket is
given by
\begin{equation}
|P(t)\rangle=\sum_{\{s_i\}}P(\{s_i\};t)|\{s_i\}\rangle ,
\end{equation}
and the equation governing the time evolution is
\begin{equation}
\partial_t|P(t)\rangle=-{\cal H}|P(t)\rangle ,
\end{equation}
where, using a representation in terms of Pauli matrices, and defining
$n_k=(1-\sigma_k^z)/2$, $v_k=1-n_k$, $s_k^{\pm}=(\sigma_k^x\pm
i\sigma_k^y)/2$, we have \cite{schutz2}
\begin{eqnarray}
& & {\cal H}={1\over 2}\sum_{k=1}^{\infty}\left(D[n_kv_{k+1}+v_kn_{k+1}-
s_k^+s_{k+1}^- - s_k^-s_{k+1}^+] \right. \nonumber \\
& & \left. \quad +2\lambda[n_kn_{k+1}-s_k^+s_{k+1}^+]
\right)+{\alpha\over 2}\sum_{k=2}^{\infty}(1-\sigma_{k-1}^x
\sigma_{k+1}^x)n_k \nonumber \\
& & \quad =DH^{\rm SEP}+\lambda H^{\rm RSA}+\alpha H^{\rm
BARW} \label{spinh} \\
& & \quad =D\sum_{k=1}^{\infty}h_k^{\rm SEP}+\lambda
\sum_{k=1}^{\infty}h_k^{\rm 
RSA}+\alpha\sum_{k=2}^{\infty}h_k^{\rm BARW} . \nonumber
\end{eqnarray}
Here we have used some of the notation of \cite{schutz2}, where SEP
(symmetric exclusion process) refers to the diffusion piece, RSA
(random--sequential adsorption) to the annihilation piece and BARW
to the branching piece of the ``quantum Hamiltonian''. Notice that 
the boundary has been included in (\ref{spinh}), since particles may
not hop to the left of site $1$, and the
annihilation/branching processes have also been restricted to sites
$1,2,3\ldots$. Hence, the above operator ${\cal H}$ governs the
evolution of a $1+1$ dimensional BARW system {\it without} an
$A\to\emptyset$ reaction at the 
boundary. Averages are calculated using the projection state
$\langle|=\sum_{\{s_i\}} \langle\{s_i\}|$, i.e. $\langle {\cal
F}\rangle = \langle | {\cal F}|P(t)\rangle$.
Following \cite{schutz2}, we now define an operator ${\cal D}$ where
\begin{equation}
{\cal D}=\gamma_{-1}\gamma_0\gamma_1\gamma_2\ldots ,
\end{equation}
with
\begin{eqnarray}
& & \gamma_{2k-1}={1\over 2}[(1+i)\sigma_k^z-(1-i)] , \nonumber \\
& & \gamma_{2k}={1\over 2}[(1+i)\sigma_k^x\sigma_{k+1}^x-(1-i)] .
\end{eqnarray}
Defining a new ``quantum Hamiltonian'' as $\tilde{\cal
H}=[{\cal D}^{-1}H{\cal D}]^T$, we find
\begin{eqnarray}
& & \tilde{\cal H}=[D-\lambda]\sum_{k=1}^{\infty}h_k^{\rm
BARW}+[\alpha+\lambda]\sum_{k=1}^{\infty}h_k^{\rm SEP}+\lambda
\sum_{k=1}^{\infty}h_k^{\rm RSA} \nonumber \\
& & \qquad\qquad\qquad+{\lambda\over
2}[n_1n_0-s_1^+s_0^++n_1v_0-s_1^+s_0^-] ,
\end{eqnarray}
where we have used the commutation rules described in detail in
\cite{schutz2}. Hence, when $D=\lambda+\alpha$, we have the following
processes occurring:  
\begin{eqnarray}
& \emptyset_i A_{i+1} \leftrightarrow A_i \emptyset_{i+1} & ~{\rm
rate} ~(\lambda+\alpha)/2, \nonumber \\
& & \qquad\qquad~~~~\; i=1,2,3,\ldots, \nonumber \\
& A_i A_{i+1} \to \emptyset_i\emptyset_{i+1} & ~{\rm
rate}~\lambda,\qquad\:i=1,2,3,\ldots, \nonumber \\
& \emptyset_{i-1} A_i \emptyset_{i+1} \leftrightarrow A_{i-1} A_i
A_{i+1} 
& ~{\rm rate}~\alpha/2,\,~~\:\, i=1,2,3,\ldots, \nonumber \\
& \emptyset_{i-1}A_iA_{i+1}\leftrightarrow A_{i-1}A_i\emptyset_{i+1}
& ~ {\rm rate}~\alpha/2,\,~~\;\,i=1,2,3,\ldots, \nonumber \\
& A_0 A_1 \to \emptyset_0\emptyset_1 & ~{\rm rate}~\lambda/2,
\nonumber \\
& \emptyset_0 A_1 \to A_0 \emptyset_1 & ~{\rm rate}~\lambda/2 .
\label{edgeprocess}  
\end{eqnarray} 
Excepting the boundary terms, we see that the Hamiltonian has been
mapped back onto itself. Furthermore, at the edge, the particles may
only hop from site $1$ to site $0$
but never the other way round. This means that we can forget
about the $0$-th site in exchange for allowing the
processes $A_1 A_2\leftrightarrow A_1 \emptyset_2$ (with rate
$\alpha/2$), and $A_1\to\emptyset_1$
(with rate $\lambda/2$). Hence, we see that the new Hamiltonian
$\tilde{\cal H}$ corresponds to the case where $\mu_s\neq 0$, with the
DP processes $A\leftrightarrow A+A$ and $A\to\emptyset$ generated on
the boundary. 

If we choose the initial condition to be an uncorrelated state
with density $1/2$, denoted by $|1/2\rangle$, then the density at site
$k$, $\rho_k(t)$, is given by
\begin{equation}
\label{exeq}
\rho_k(t)=\langle|n_k\exp(-{\cal H}t)|1/2\rangle .
\end{equation}
Following exactly the procedure in \cite{schutz2,schutz1} (starting
with insertions of the identity operator ${\cal D}{\cal D}^{-1}$ into
the RHS of (\ref{exeq})), one can straightforwardly show that
\begin{equation}
\label{equiv}
\rho_k(t)={1\over 2}[1-\langle 0|\exp(-\tilde{\cal H}t)|k-1,k\rangle] ,
\end{equation}
where $\langle 0|$ is the vacuum state (with no particles), and
$|k-1,k\rangle$ is the initial state with only two particles situated
at sites $k-1$ and $k$. However the relation (\ref{equiv}) is just
what we wanted to prove: the LHS is the density at the $k$-th site,
whereas the RHS is one-half times the probability that a cluster
initiated at $t=0$ by two particles
at sites $k-1$ and $k$ has not yet died out by time
$t$. According to our earlier analysis, for $\Delta<0$, the LHS should
scale as 
$|\Delta|^{\beta_{\rm dens}}$ (far from the wall) or
$|\Delta|^{\beta^{\rm Sp}_{1,{\rm dens}}}$ (close to the wall), and
the RHS as $|\Delta|^{\beta_{\rm seed}}$ (far from the wall) or
$|\Delta|^{\beta^{\rm O}_{1,{\rm seed}}}$ (close to the wall). Thus,
at the line $D=\lambda+\alpha$, we have shown the desired result
$\beta^{\rm O}_{1,{\rm seed}}=\beta^{\rm Sp}_{1,{\rm dens}}$ (and, of
course, the bulk result $\beta_{\rm seed}=\beta_{\rm dens}$). We note
that the bulk result was proven in \cite{schutz2}, and a very similar
result for $A+A\to\emptyset$ was derived in \cite{schutz1} (connecting
the O* and Sp* ``transitions'').
Using universality, we postulate that the equality between the two
surface exponents is valid everywhere close to the transition line,
and not just where $D=\lambda+\alpha$. 

It is now straightforward to derive the relation
$\beta^{\rm Sp}_{1,{\rm seed}}=\beta^{\rm O}_{1,{\rm
dens}}$ (again at the line $D=\lambda+\alpha$). One simply starts off
with the quantum Hamiltonian $\tilde{\cal H}$ and then
follows the same steps as above. $\tilde{\cal H}$ can then be mapped
back onto the starting Hamiltonian ${\cal H}$, meaning that the
transformation is actually a duality transformation. A relation
like that in (\ref{equiv}) can then be derived, giving
$\beta^{\rm Sp}_{1,{\rm seed}}=\beta^{\rm O}_{1,{\rm
dens}}$. 

In summary, at the particular line in parameter space
$D=\lambda+\alpha$, we have derived some exponent equalities which are
in full agreement with the simulations to be presented in the next
section. In particular, we see that we have mapped BARW at the
special transition onto BARW at the ordinary transition in $1+1$
dimensions (and vice-versa), a
rather non-trivial procedure. This has allowed us to derive some
results about the $\beta_1$ exponents (something which seems to
be beyond the ability of the field theoretic methods at
present). Unfortunately, as is always the case with exact calculations,
the result is only derived for one line in parameter space and we have
to rely on universality in order to claim that it is valid
elsewhere close to the transition line. 

\section{DP\lowercase{$n$} and BARW models}
\label{dpnbarw}

We will now briefly present the specific models and boundary
conditions used in our numerical simulations. We begin with
DP, include its generalization to DP2, and then comment on how we
implement BARW. In all cases we include the specific boundary
conditions and identify them according to the classification in
sections~\ref{DPsurface} and~\ref{BARWsurface}.

\begin{figure}
\begin{center}
\leavevmode
\vbox{
\epsfxsize=3in
\epsffile{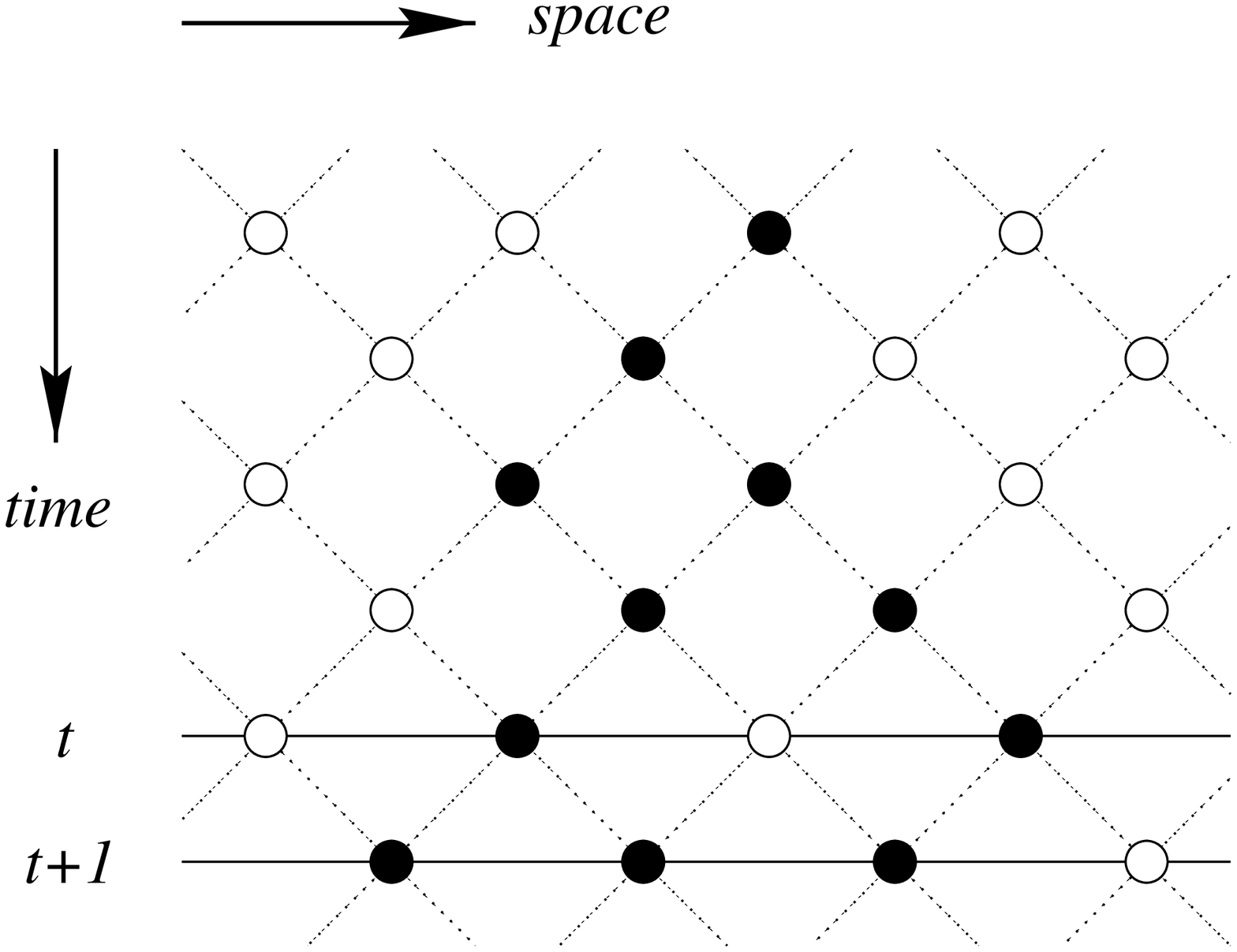}}
\end{center}
\caption{Directed Percolation in terms of the Domany-Kinzel model,
where time flows vertically downwards. Black sites are
active ($A$) and white ones inactive ($I$). The state of each site at
time $t+1$ depends on the states of the neighboring sites at time $t$.}
\label{DPlattice}
\end{figure}

\begin{figure}
\begin{center}
\leavevmode
\vbox{
\epsfxsize=2.5in
\epsffile{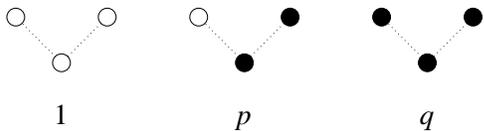}}
\end{center}
\caption{Update probabilities for DP in terms of the parameters
$0 \leq p, q \leq 1$, where we have $q=p(2-p)$ for bond DP
and $q=p$ for site DP, respectively. Probabilities for the other
configurations follow from left-right symmetry and from
$P(A \mid \ldots) + P(I \mid \ldots) = 1$.}
\label{DPrules}
\end{figure}

For $d=1$, bond DP as well as site DP (for which the sites
percolate instead of the bonds), are contained in the Domany-Kinzel
model \cite{domany1,domany2}.
Each site can either be active or inactive and the probability
for site $i$ to be updated to state $s_{i,t+1}$ at time $t+1$ is
given by $P(s_{i,t+1} | s_{i-1,t}, s_{i+1,t})$. See
Figure~\ref{DPlattice} for a typical lattice configuration and
Figure~\ref{DPrules} for the update rules.

\begin{figure}[htb]
\centerline{\hbox{
\epsfxsize=1.5in
\epsfbox{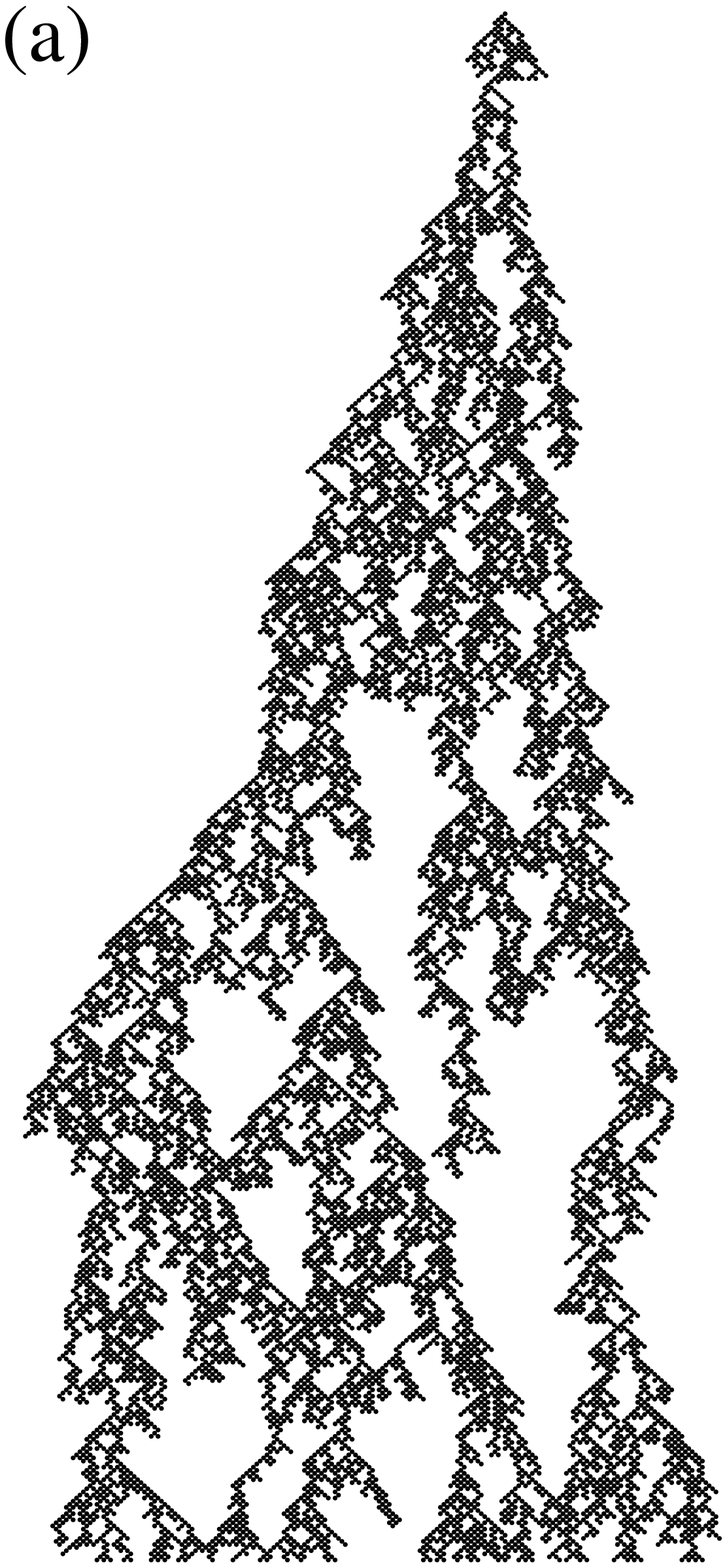}}
\epsfxsize=1.5in
\epsfbox{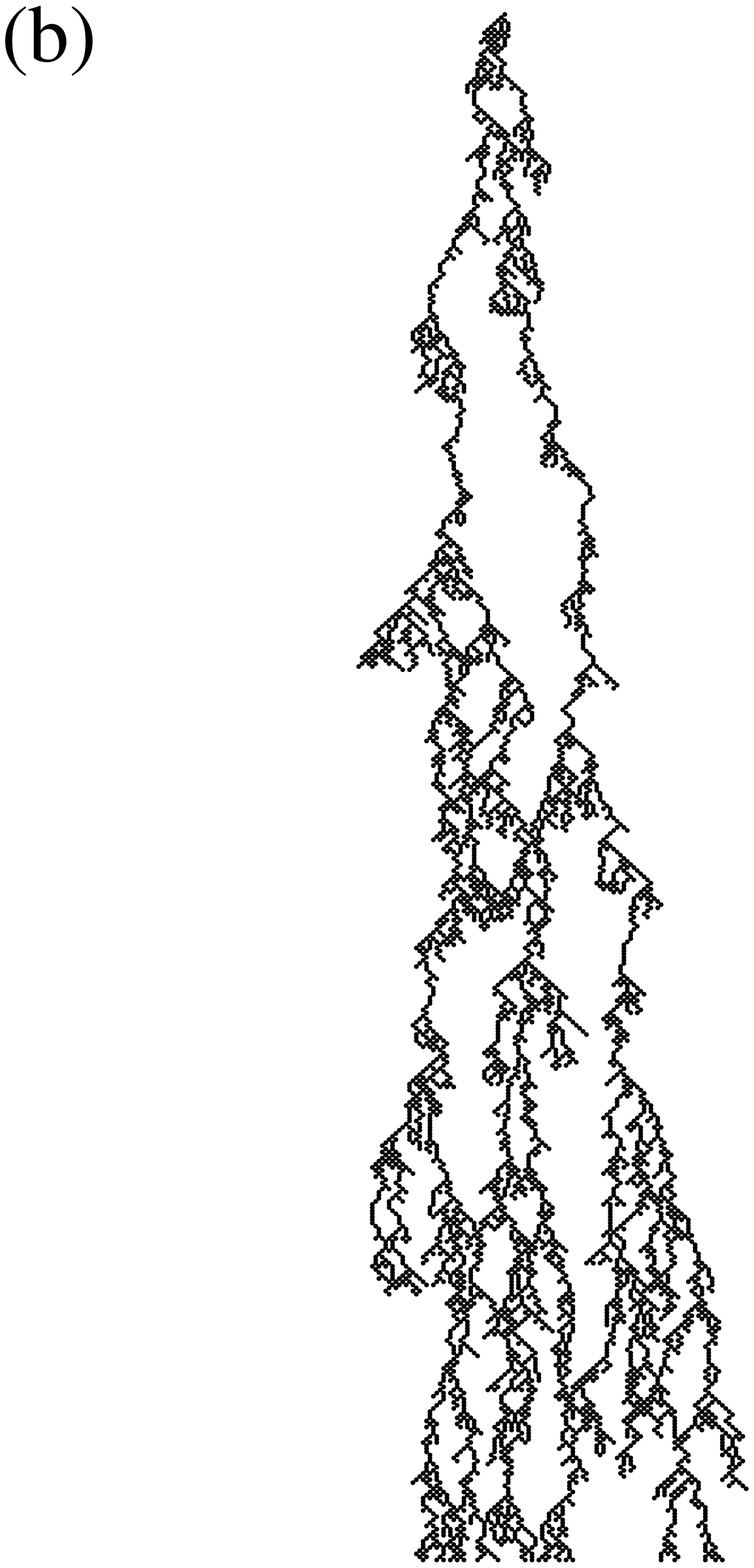}}
\vspace*{5mm}
\caption{(a) A DP cluster and (b) a DP2 cluster both grown from
a single seed in the bulk.} 
\label{dp2fig}
\end{figure}

\begin{figure}
\begin{center}
\leavevmode
\vbox{
\epsfxsize=3in
\epsffile{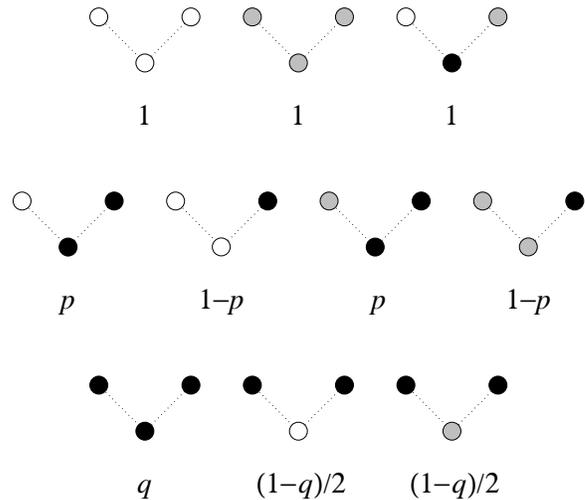}}
\end{center}
\caption{Update probabilities for DP2: black sites are active ($A$),
whereas 
white and grey sites are in the inactive states $I_1$ and $I_2$,
respectively.  Probabilities for the other configurations follow from
left-right symmetry and from $P(A \mid \ldots) + P(I_1 \mid \ldots)
+ P(I_2 \mid \ldots) = 1$.}
\label{DP2rules}
\end{figure}

The DP2 model has two symmetric absorbing states in which the system
can be trapped. It is a special case of a generalized Domany-Kinzel model
(DP$n$) introduced by Hinrichsen \cite{haye}, where each site can be
either active 
or in one of $n$ inactive states. For $n=1$ the update rules are
identical to those of the Domany-Kinzel model in Figure~\ref{DPrules},
but for $n \geq 2$, the distinction between regions of different inactive
states is preserved by demanding that they are separated by active
ones. An example of a DP2 cluster is shown in Figure~\ref{dp2fig}b,
where we have also shown an ordinary DP cluster in
Figure~\ref{dp2fig}a for comparison. In $1+1$ dimensions, DP2 belongs
to the BARW universality class, 
and the update probabilities are given in Figure~\ref{DP2rules}.

\begin{figure}
\begin{center}
\leavevmode
\vbox{
\epsfxsize=3in
\epsffile{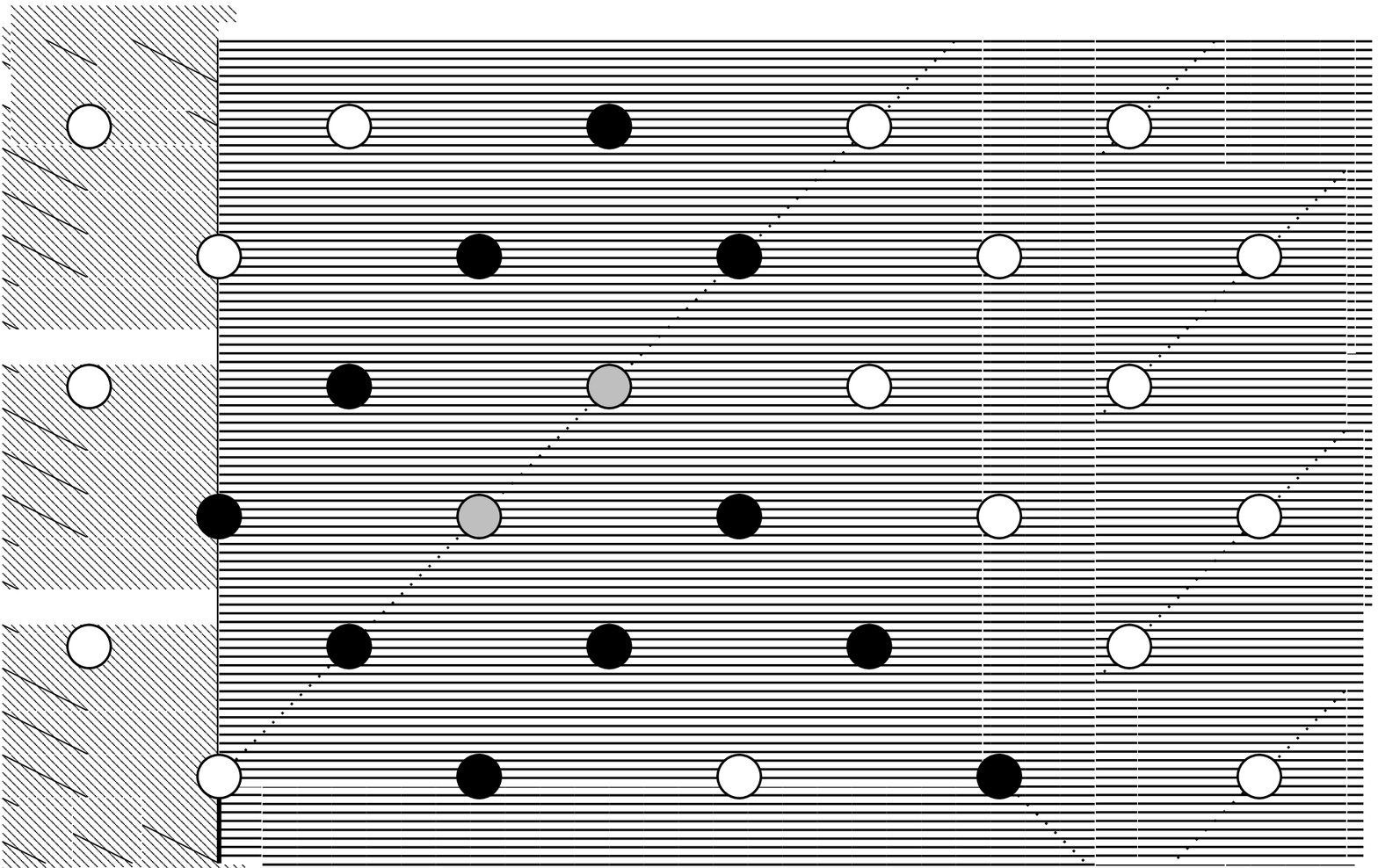}}
\end{center}
\caption{DP2 with an inactive boundary condition (IBC), corresponding
to the special (Sp) universality class.}
\label{DP2_IBC}
\end{figure}

The easiest way of introducing a boundary into DP and DP2
is simply to cut off the lattice. This is equivalent to
introducing boundary sites which are forced to be in one of the
inactive states. We will refer to this case as the inactive
boundary condition (IBC) and we choose
inactivity of type $1$ to the left of the boundary, 
see Figure~\ref{DP2_IBC}. Apart from imposing the state of these sites
within the wall, the sites at the wall and those in the bulk are
updated by the rules in Figures~\ref{DPrules} and~\ref{DP2rules}.

\begin{figure}
\begin{center}
\leavevmode
\vbox{
\epsfxsize=3in
\epsffile{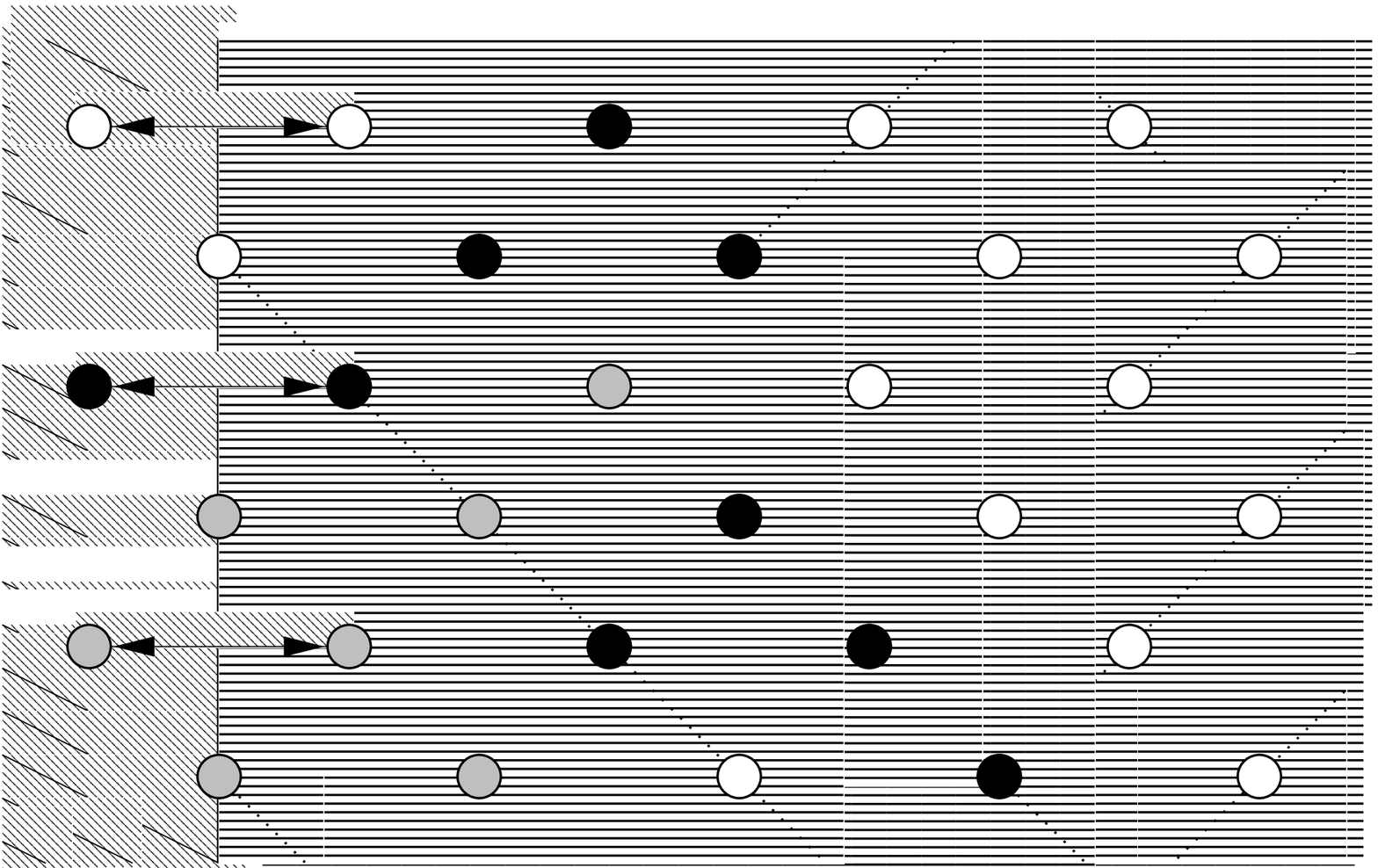}}
\end{center}
\caption{DP2 with a reflecting boundary condition (RBC), corresponding
to the ordinary (O) universality class.}
\label{DP2_RBC}
\end{figure}

Next we consider the reflecting boundary condition (RBC) where the
wall acts like a mirror so that the sites within the wall
are always a mirror image of those next to the wall, see
Figure~\ref{DP2_RBC}. For DP2, one can see that there is a qualitative
difference between the IBC and the RBC. For the latter, regions of
type-$2$ inactivity can get trapped at the wall and the only way for
these regions to disappear is to wait
for the cluster to return, whereas for the IBC such regions are never
trapped. 

\begin{figure}
\begin{center}
\leavevmode
\vbox{
\epsfxsize=3in
\epsffile{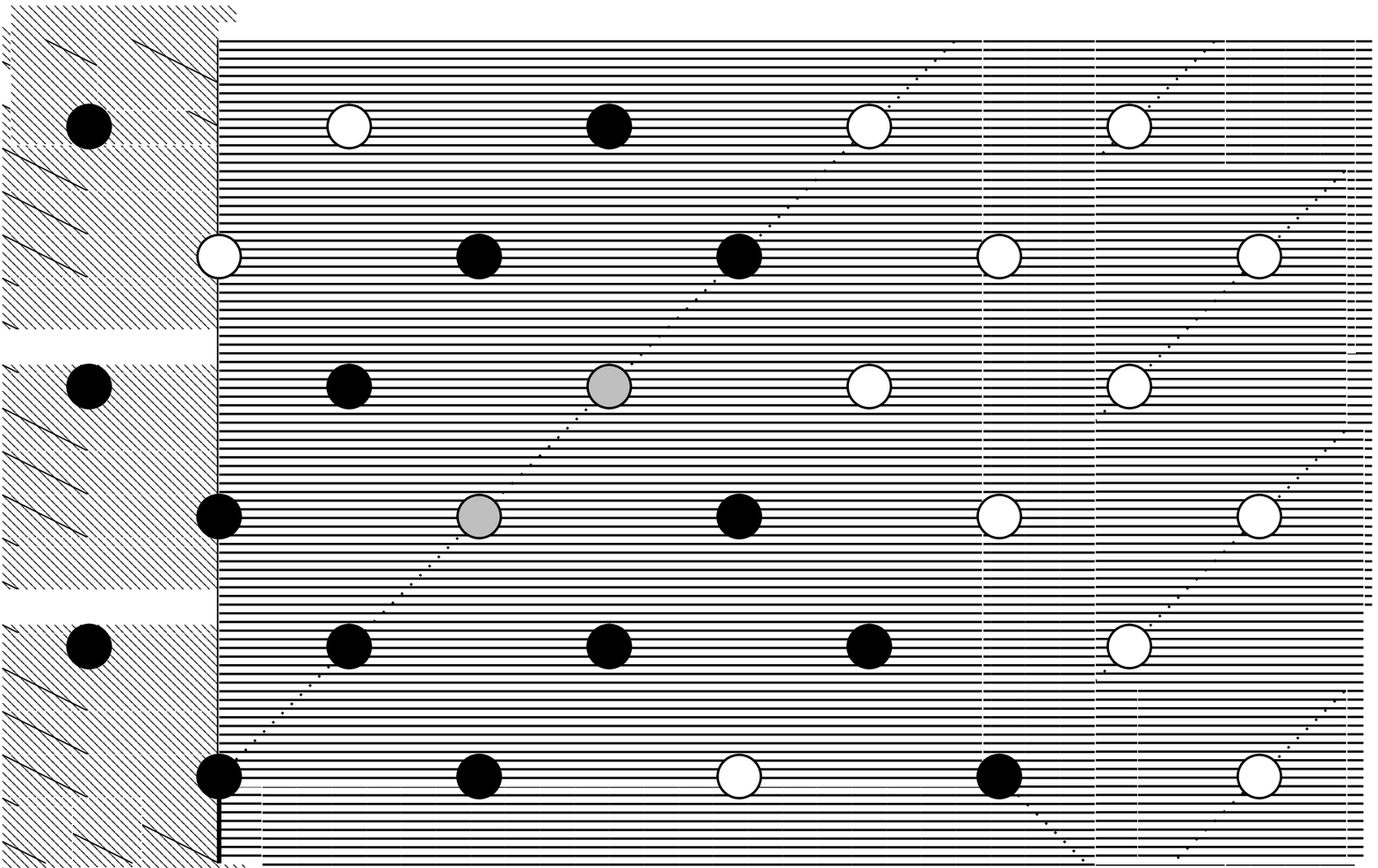}}
\end{center}
\caption{DP2 with an active boundary condition (ABC), corresponding
to the extraordinary (E) universality class.}
\label{DP2_ABC}
\end{figure}

We now consider the active boundary condition (ABC) where the
sites within the wall are forced to be active, see
Figure~\ref{DP2_ABC}. 
In this case the cluster will never die completely as the wall will
always be active and can always induce new clusters. Nevertheless, by 
defining the survival time of a cluster as the point in time when the
system has no activity apart from within the wall itself, we can
define the same exponents for the ABC as for the other boundary
conditions.  We have, however, not studied this boundary condition in
any detail but merely mention it here for completeness. 

We can now discuss the relation between the above boundary conditions
and our previous classification of the universality classes at the
boundary 
for BARW in $1+1$ dimensions. The key feature is whether the
symmetry between the two absorbing states in the bulk is preserved at
the surface. In terms
of the DP2 model, the IBC model respects this symmetry and hence it
belongs to the special (Sp) universality class, whereas the RBC
model 
does not respect this symmetry, and hence belongs to the ordinary (O)
universality class. Furthermore, the ABC model clearly belongs
to the extraordinary (E) universality class. Hence we see that by
using the IBC, 
RBC, ABC classification all the previously discussed boundary BARW 
transitions in $1+1$ dimensions can be accessed. 

Furthermore, let us note that for DP the classification of the IBC
and RBC is somewhat different. In $1+1$ dimensions DP probably
does not support the special transition, and since no symmetry
is broken by the RBC, both the IBC and RBC will belong to the ordinary
transition universality class \cite{dp-wall-edge,ourprl}.

We have also performed simulations for a lattice BARW model with IBC
and RBC boundary conditions. For BARW we initially placed two
particles at the two sites closest to the wall. The one-dimensional
BARW model is then implemented with ``dynamic branching'', 
which means that the branching of one particle
into three particles occurs randomly to either the left or
the right of the particle \cite{kwon-park:1995}.
The BARW model is expected to be in the same universality class
as DP2 and our results are in agreement with this
for both bulk and surface quantities (using the IBC and RBC).
Hence, in the following section, we will only discuss the results for
DP2 since this is also the model to which we devoted most of our
simulations.

\section{Numerical results}
\label{numres}

We have studied DP2 in detail using Monte Carlo simulations in $1+1$
dimensions. The wall is placed at $i=0$ and we use an 
initial configuration with one active site at $i=0$,
with the sites $i>0$ being in the inactive state $I_1$.
Thus the absorbing state corresponds to the situation where all
sites are in the inactive state $I_1$.
The system is evolved according to the DP2 rules (see
Figure~\ref{DP2rules}), and we typically average
over $10^5$ independent clusters in order to reduce the error bars
to a few percent. Using the notation of Figure~\ref{DP2rules},
we have carried out simulations for $q=p$ at the critical
probability $p_c$, where we have used the estimate $p_c=0.5673$
\cite{haye}. 

In these simulations, starting from a seed on the wall,
we measure the survival probability $P_1(t)$, the activity
in the bulk $N_1(t)$ and at the wall $N_{1,1}(t)$, the average
spread of the cluster $\left< x^2(t) \right>$,
and the probability $p_1(s)$ to have a cluster of
size (mass) $s$, all at criticality \cite{haye,grassberger-torre}.
Furthermore, by averaging over surviving clusters only (denoted by an
over--line), we measure the surviving bulk activity
$\overline{N_{1}}(t)$ 
and the surviving wall activity $\overline{N_{1,1}}(t)$,
again starting from a seed on the wall.

First we performed simulations for DP2 without a wall and obtained
results for the exponents in complete agreement with those in
\cite{haye}.
Our results are listed in Tables \ref{table-exp1} and \ref{table-exp2}.
There are several estimates available for the bulk exponent
$\beta_{\rm dens}$ ($= \beta_{\rm seed}$) 
\cite{jensen:1997}. In the following we will use the estimate
$\beta_{\rm dens}=0.922(5)$ \cite{zhong}. 

We now list some exponent relations used to extract the exponents
from our numerical simulations \cite{exponen}. All the
relations given below are valid for both 
the IBC (special) and RBC (ordinary) DP2 transitions, and hence
these labels will be suppressed from now on.
The probability for a cluster grown from a seed on the wall
still to be alive at time $t$ is given by Eq.~(\ref{barwsurvwall}). At
criticality ($\Delta = 0$) it has the following behavior
\begin{equation}
        P_1(t) \sim t^{-\delta_{1, \rm seed}} ,
                                          \label{eq:P_1(t)}
\end{equation}
with the exponent
\begin{equation}
        \delta_{1, \rm seed} = \beta_{1, \rm seed}/\nu_{\parallel} .
                                          \label{eq:delta_1,seed}
\end{equation}
Hence, the probability of growing a cluster which lives exactly $t$
time steps behaves as $p_1(t)\sim t^{-1 - \delta_{1, \rm seed}}$.
Away from criticality it is straightforward to obtain the average
cluster lifetime of finite clusters from (\ref{barwsurvwall}). One
obtains 
\begin{equation}
        \left< t \right> \sim |\Delta|^{-\tau_{1}},
                                          \label{eq:<t>}
\end{equation}
with the exponent
\begin{equation}
        \tau_{1} = \nu_\parallel - \beta_{1, \rm seed} .
                                          \label{eq:tau_1}
\end{equation}

The average number of active sites at criticality,
averaged over all clusters, is obtained by integrating 
the density (\ref{ansatz_rho_wall}) over space,
and one arrives at
\begin{equation}
        N_1(t) \sim t^{\kappa_1},
                                        \label{eq:N_1(t)}
\end{equation}
with
\begin{equation}
        \kappa_1 = d\chi - \delta_{\rm dens} - \delta_{1, \rm seed}, 
                                        \label{eq:kappa_1}
\end{equation}
where we have introduced the cluster envelope or ``roughness'' exponent
$\chi=\nu_\perp/\nu_\parallel$ ($\equiv 1/z$), and the notation
$\delta_{\rm dens} = \beta_{\rm dens}/\nu_{\parallel}$.
Note that (\ref{eq:kappa_1}) corresponds to the hyperscaling relation
(\ref{surf_hyperscaling}), a fact which follows from
$\langle s_1(t) \rangle = \int_0^t dt' \, N_{1}(t')$, and the relation
$\gamma_1 = \nu_\parallel (1+\kappa_1)$.
By integrating the density on the wall (\ref{ansatz_rho_wall_wall}),
we obtain the average number of active sites at criticality
on the wall
\begin{equation}
        N_{1,1}(t) \sim t^{\kappa_{1,1}},
                                        \label{eq:N_1,1(t)}
\end{equation}
with
\begin{equation}
        \kappa_{1,1} 
             = (d-1)\chi - \delta_{1, \rm dens} - \delta_{1, \rm seed}, 
                                        \label{eq:kappa_1,1}
\end{equation}
and where
$\delta_{1, \rm dens} = \beta_{1, \rm dens}/\nu_{\parallel}$.
Note also that (\ref{eq:kappa_1,1}) corresponds to the hyperscaling
relation (\ref{surf_hyperscaling_wall_wall}) at criticality,
since $\langle s_{1,1}(t) \rangle = \int_0^t dt' \, N_{1,1}(t')$, and 
$\gamma_{1,1} = \nu_\parallel (1+\kappa_{1,1})$.

Alternatively, by averaging only over clusters
which survive to infinity (denoted by an over--line), we obtain
\begin{equation}
        \overline{N_{1}}(t) \sim t^{\overline{\kappa_1}},
                                        \label{eq:overline-N_1}
\end{equation}
where 
\begin{equation}
        \overline{\kappa_1} = d\chi - \delta_{\rm dens}.
                                        \label{eq:overline-kappa-1}
\end{equation}
The activity on the wall for surviving clusters reads
\begin{equation}
        \overline{N_{1,1}}(t) \sim t^{\overline{\kappa_{1,1}}} ,
                                        \label{eq:overline-N_1,1}
\end{equation}
with the exponent
\begin{equation}
        \overline{\kappa_{1,1}} = (d-1)\chi - \delta_{1, \rm dens}.
                                        \label{eq:overline-kappa_1,1}
\end{equation}
Simulations in $1+1$ dimensions thus directly yield
$\delta_{1, \rm dens}$.
The average position of activity follows from (\ref{ansatz_rho_wall})
\begin{equation}
        \left< x^2 \right> \sim  t^{2\chi} ,
                                        \label{eq:<x^2>}
\end{equation}
where $x$ is the distance from the seed and the average is taken
over all active points at a given time.

For further confirmation of our numerical data we also considered the
cluster size distributions at criticality. In the bulk the typical
cluster size $s$ of finite clusters scales as volume times density,
i.e. 
\begin{equation}
        s \sim \xi_\perp^d \xi_\parallel n(\Delta) \sim
        |\Delta|^{-1/\sigma}, 
                                \label{eq:s-typical}
\end{equation}
with 
\begin{equation}
        1/\sigma = d\nu_\perp + \nu_\parallel - \beta_{\rm dens}.
                                \label{eq:1/sigma}
\end{equation}
{}From the lifetime survival distribution (\ref{eq:P_1(t)}), it is
then straightforward to obtain the probability $P_1(s)$ to have a
cluster of size larger than $s$, for clusters started from a seed on
the wall. Using the fact that the lifetime
is set by the parallel correlation length,
$t\sim \xi_\parallel \sim |\Delta|^{-\nu_\parallel}$, we see that
the typical cluster size and lifetime are connected by
\begin{equation}
        s \sim t^{1/\nu_\parallel\sigma} .
\end{equation}
Hence we obtain $P_1(s) \sim P_1(t\sim s^{\nu_\parallel\sigma})
\sim s^{-\beta_{1,{\rm seed}} \sigma}$.
Thus, we eventually obtain the probability $p_1(s)$ to have
a cluster of exactly size $s$, $p_1(s)= - dP_1(s)/ds$, with the
result
\begin{equation}
   p_1(s) \sim s^{-\mu_1},
                                \label{eq:p_1(s)}
\end{equation}
where
\begin{equation}
   \mu_1 = 1 + \frac{\beta_{1, \rm seed}}
               {d \nu_\perp + \nu_\parallel - \beta_{\rm dens}}.
                                \label{eq:mu_1}
\end{equation}
Similarly, the cluster size distribution on the wall due to a 
seed located at the wall can also be obtained.
In this case the typical cluster size of finite clusters is
\begin{equation}
        s_{\rm wall} \sim \xi_\perp^{d-1} \xi_\parallel n_1(\Delta)
        \sim |\Delta|^{-1/\sigma_1} ,
                                \label{eq:s_wall-typical}
\end{equation}
where
\begin{equation}
        1/\sigma_1 = (d-1)\nu_\perp + \nu_\parallel - \beta_{1, \rm
        dens} . 
                                \label{eq:1/sigma_1}
\end{equation}
The resulting distribution reads
\begin{equation}
        p_{1,1}(s_{\rm wall}) \sim s_{\rm wall}^{-\mu_{1,1}},
                                \label{eq:p_1,1(s)}
\end{equation}
with
\begin{equation}
        \mu_{1,1} = 1 + \frac{\beta_{1, \rm seed}}
             {(d-1) \nu_\perp + \nu_\parallel - \beta_{1, \rm dens}}.
                                \label{eq:mu_1,1}
\end{equation}
Note also that many of the scaling expressions given above only apply
exactly at bulk criticality. Away from that point one must also include a
scaling function. For example,
Eq.~(\ref{eq:P_1(t)}) is replaced by
$P_1(t,\Delta) = |\Delta|^{\beta_{1, \rm seed}} \,
F(t/|\Delta|^{-\nu_{\parallel}})$, and
Eq.~(\ref{eq:p_1(s)}) is replaced by
$p_{1}(s,\Delta) = |\Delta|^{\mu_1/\sigma} \,
G(s/|\Delta|^{-1/\sigma})$. 

In Tables \ref{table-exp1} and \ref{table-exp2}
we list our estimates for the critical exponents 
for DP2. Our results are in
complete accordance with our theoretical analysis: bulk exponents are
unaltered whereas the wall introduces two separate surface exponents.
We have also carried out bulk and surface simulations for $\Delta<0$
and confirmed that our data can be collapsed according to an
appropriate survival probability scaling function [see
(\ref{barwsurvwall}) for the surface case], using our exponent
estimates. This numerically confirms the validity of the relation
$\delta=\beta/\nu_\parallel$ for the bulk, as well as the analogous
relations for the both sets of
surface exponents \cite{no-theory-for-it}. 

By using the explicit definitions of the IBC and RBC, we can deduce
some further properties of the $\beta_{1, \rm seed}$ and $\beta_{1,
\rm dens}$ exponents. There will be more activity next
to the wall for the IBC than for the RBC, since 
the latter can have regions of $I_2$ located at the wall.
Once created, these $I_2$ regions will survive until the
activity returns to the wall. Thus it follows 
that $\beta_{1, \rm dens}^{\rm IBC} \leq \beta_{1, \rm dens}^{\rm RBC}$.
On the other hand, the existence of these $I_2$ regions implies that
the survival probability (\ref{barwsurvwall})
for the RBC will be greater than for the IBC, leading 
to $\beta_{1, \rm seed}^{\rm IBC} \geq \beta_{1, \rm seed}^{\rm RBC}$.
Note that both our simulations and our previous exact calculations
show that $\beta_{1,{\rm seed}}+\beta_{1,{\rm dens}}$ is the same for
both the RBC and IBC. Using a hyperscaling relation [like that in
(\ref{surf_hyperscaling_wall_wall})], this
implies that $\gamma_{1,1}^{\rm IBC}=\gamma_{1,1}^{\rm RBC}$ (although
both exponents defined in this way are negative).
We have also studied several other boundary conditions and found that
these give the 
same scaling behavior as either the RBC or IBC depending on whether
the above-mentioned $I_2$ regions can disappear only at the wall
or also in the bulk. 

We can obtain an interesting exponent relation for the RBC transition
by assuming that the survival probability is dominated by the return
to the wall of the cluster-envelope, which leads to \cite{lauritsen-etal} 
\begin{equation}
	\delta^{\rm RBC}_{1, \rm seed} = 1-\chi ,
				\label{eq:1-chi}
\end{equation}
in agreement with our simulation results for the RBC. 
Qualitatively, this means that the $I_2$ regions located at the wall
determine the scaling since they can only disappear when the activity
returns to the wall. Note that a relation of this kind clearly fails
for the IBC transition. Furthermore, if the cluster lifetime is
defined to be the return time of the cluster-envelope (i.e.\ the return
time of the rightmost active site) to the initial point, then
we expect clusters defined in this way to have a lifetime distribution 
exponent $\delta_{1,{\rm seed}}$ given by Eq.~(\ref{eq:1-chi}). This
prediction is in agreement
with the simulations in \cite{hwang-etal:1998},
where various models in the DP and BARW classes were studied
with cluster lifetimes defined in the way described above.

For DP it has been customary for some time to investigate
whether the critical exponents can be fitted by simple rational
numbers \cite{jensen:1996}. Such a fitting has also been tried for 
bulk BARW with the following guesses in $1+1$ dimensions:
$\kappa = \chi - 2\delta = 0$ and $\chi=4/7$ \cite{jensen:1994}. 
These estimates lead immediately to $\delta=2/7$
(and $\beta/\nu_\perp=1/2$, $\gamma=\nu_\parallel$).
It is intriguing to note that our numerical results for DP2
in addition suggest that $\mu_{1} = 3/2$ for the IBC and $4/3$ for the
RBC. From Eq.\ (\ref{eq:mu_1}), it then follows that 
$\delta_{1, \rm seed} = 9/14$ for the IBC and $3/7$ for the RBC.
We would need one more relation in order to obtain the last 
independent exponent, which we can take to be $\nu_\parallel$.
In fact, we observe numerically that the relation 
$2\nu_\parallel-\beta_{1, \rm seed}-\beta_{1, \rm dens} = 3$
is valid to within one percent.
{}From these observations the remaining DP2 exponents follow:
$\beta_{\rm dens} = \beta_{\rm seed} = 12/13$,
$\nu_\parallel=42/13$ and $\nu_\perp=24/13$. Furthermore, 
$\beta_{1, \rm seed}=27/13$ and $\beta_{1, \rm dens}=18/13$ for the
IBC, and vice versa for the RBC. However, at present we have
no understanding of these possible exact values for the
$1+1$ dimensional exponents. BARW is certainly not conformally
invariant and 
consequently until some theoretical framework is proposed to explain
why these exponents could be rational numbers, numerical coincidence
remains a distinct possibility.  

\section{Conclusion}
\label{conc}

In this paper we have presented a study of critical surface
effects in systems with non-equilibrium phase transitions. In
particular we have focused on the DP and BARW universality classes,
where we have put forward a unified presentation involving mean-field,
scaling, field-theoretic and exact methods. Furthermore, many of our
theoretical conclusions have been backed up by large-scale
Monte-Carlo simulations. 

Nevertheless, there are still a number of open questions. In particular
our understanding of surface BARW in $1+1$ dimensions is hampered
by the fundamental problems of the field theory, which mean that the
boundary (and bulk) transitions occurring at $\sigma_m=\sigma_{m,{\rm
critical}}$ remain difficult to treat using RG methods. Furthermore,
we have concentrated on the ordinary and special transitions in $1+1$ 
dimensional BARW---there will most likely also be interesting 
behavior for the extraordinary transition. 

Finally, our most important result is the existence of two independent
surface exponents: $\beta_{1,{\rm dens}}$ and $\beta_{1,{\rm seed}}$
for surface BARW (and DP2). 
This certainly distinguishes DP from BARW, since for the
former case, $\beta_{1,{\rm seed}}=\beta_{1,{\rm dens}}$. For the
$1+1$ dimensional BARW
case on the other hand, we have used exact techniques to link the
surface exponents at the ordinary and special transitions, giving
$\beta^{\rm O}_{1,{\rm seed}}=\beta^{\rm Sp}_{1,{\rm dens}}$ and
$\beta^{\rm Sp}_{1,{\rm seed}}=\beta^{\rm O}_{1,{\rm dens}}$. It would
certainly be instructive to rederive these results from a
field theoretic perspective, but this is beyond the scope of the
present paper.

\section*{acknowledgments}

We would like to thank Tim Newman, Beate Schmittmann, and Uwe T\"auber
for very useful discussions. M. H. acknowledges support from the US
National Science Foundation through the Division of Materials
Research, and is grateful for hospitality and financial support from
the CATS group at the Niels Bohr Institute, where part of this work
was performed. P. F. acknowledges support from the Swedish Natural
Science Research Council. K. B. L. acknowledges support from the
Carlsberg Foundation. 

\end{multicols}

\bigskip
\widetext
\begin{multicols}{2}

\end{multicols}

\widetext
\begin{table}[htb]
\caption{Critical exponents obtained from our DP2 simulations.
        For comparison we also list the exponents for DP in the bulk and
        with an IBC wall
        \protect\cite{lauritsen-etal,essam-etal:1996,jensen:1996}.
         The $\delta_{1,\rm seed}$ exponent is obtained from
         Eqs.~(\protect\ref{eq:N_1(t)}) and (\protect\ref{eq:kappa_1}).
         Exponents without the `1' subscript refer to the bulk.
        }
\vspace*{0.5cm}
\begin{tabular}{|c||c|c|c|c|c|} \hline
              &    DP       &  DP (IBC)  &  DP2   &  DP2 (IBC)  &  DP2 (RBC) \\
\hline\hline
$\overline{\kappa_1}$
              & 0.473 14(3) & 0.473 14(3) & 0.288(5) &  0.287(2) & 0.285(2)\\
$\delta_{\rm dens}$
              & 0.159 47(3) & 0.159 47(3) & 0.287(5) &  0.288(2) & 0.291(4)\\
$\beta_{\rm dens}$
              & 0.276 49(4) & 0.276 49(4) & 0.922(5) &  0.93(1)  & 0.94(2) \\
\hline
$\kappa$      & 0.313 68(4) &            & 0.000(2) &           &        \\
$\kappa_1$    &             & 0.0496(3)  &         & -0.354(2)~ & -0.141(2)~\\
\hline
$\delta_{1,\rm seed}$
              &             & 0.4235(3)  &           &  0.641(2) & 0.426(3) \\
$\beta_{1,\rm seed}$
              &             & 0.7338(1)  &           &  2.06(2)  & 1.37(2)  \\
\hline
$\delta_{1, \rm dens}$
              &             & 0.4235(3)  &           &  0.415(3) & 0.635(2) \\
$\beta_{1, \rm dens}$
              &             & 0.7338(1)  &           &  1.34(2)  & 2.04(2)  \\
\hline
$2\chi$        & 1.265 23(2) &           &  1.150(5) &   1.150(3) & 1.152(3)\\
$\chi$        & 0.632 61(2) &           &  0.575(3) &            & \\
$\nu_\parallel$& 1.733 83(3) &           &  3.22(3)  &            & \\
$\nu_\perp$    & 1.096 84(2) &           &  1.84(2)  &            & \\
\hline
\end{tabular}
\vspace*{0.5cm}
\label{table-exp1}
\end{table}

\begin{table}[htb]
\caption{Critical exponents for cluster lifetime (\protect\ref{eq:P_1(t)})
         and mass distributions
         (\protect\ref{eq:p_1(s)}), (\protect\ref{eq:mu_1}),
         (\protect\ref{eq:p_1,1(s)}), and (\protect\ref{eq:mu_1,1}).
         For comparison we also list the exponents for DP in the bulk
         and with an IBC wall
         \protect\cite{lauritsen-etal,essam-etal:1996,jensen:1996}.
         We also give the exponents for the average lifetime
         (\protect\ref{eq:<t>}), and average cluster sizes
         (\protect\ref{size_wall}) and
         (\protect\ref{size_wall_wall}), obtained
         from the scaling relations.
        }
\vspace*{0.5cm}
\begin{tabular}{|c||c|c|c|c|c|} \hline
              &    DP       &  DP (IBC) &  DP2      & DP2 (IBC) & DP2 (RBC)\\
\hline\hline
$\delta_{\rm seed}$
              & 0.159 47(3) &           &  0.290(5) &           &   \\
$\delta_{1, \rm seed}$
              &             & 0.4235(3) &           &  0.646(3) & 0.425(3) \\
\hline
$\mu$         & 1.108 25(2) &           &  1.225(5) &           &    \\
$\mu_1$       &             & 1.2875(2) &           &  1.500(3) & 1.336(3) \\
$\mu_{1,1}$   & 1.189 72(6) & 1.7337(2) &  1.408(5) &  2.05(5)  & 2.15(5)  \\
\hline
$\tau$        & 1.457 34(7) &           &  2.30(3)  &           &   \\
$\tau_1$      &             & 1.0002(3) &           &  1.16(4)  & 1.85(4)   \\
\hline
$\gamma$      & 2.277 69(4) &           &  3.22(5)  &           &   \\
$\gamma_1$    &             & 1.8207(4) &           &  2.08(4)  & 2.77(4)  \\
$\gamma_{1,1}$ & 1.180 85(4)& 0.2664(3) &  1.38(3)  &  ($<0$)   & ($<0$)  \\
\hline
\end{tabular}
\vspace*{0.5cm}
\label{table-exp2}
\end{table}


\begin{thebibliography}{99}
\bibitem{diffprl}
	Note that the convention used for $\Delta$ here is different
	from that in \cite{ourprl}.
\bibitem{mistake}
        Note that the mean field surface density exponent
        $\beta_{1,{\rm dens}}^{\rm IBC}$ quoted in
        footnote [22] of Ref.~\cite{ourprl} is incorrect.
\bibitem{Eexpan}
	For $\Delta>0$, we have $n_s=(3\Delta_s^2/2D\lambda)
	-(3\Delta/2\lambda)$, whereas for $\Delta<0$, we have
	$n_s=(3\Delta_s^2/2D\lambda)-(3\Delta/2\lambda)+
	(2D^2\Delta^3/9\lambda\Delta_s^4)+{\rm higher~order~
	terms}$.
\bibitem{exponen}
        Note that some of our exponent definitions in this section
        differ 
        from those sometimes found in the literature. In particular
        our $\kappa$ exponents are sometimes called $\eta$, and the
        dynamic exponent $z$ is sometimes redefined as $z\to 2/z$.
\bibitem{no-theory-for-it}
        Technical difficulties have so far prevented 
        a field-theoretic derivation of these kind of relations for BARW 
        \protect\cite{cardy-tauber}.
\end{thebibliography}
\end{document}